\newif\ifshort

\documentclass{elsarticle}

\usepackage{amsmath}
\usepackage{amssymb}
\usepackage{gastex}
\usepackage{graphicx}
\usepackage{kbordermatrix}
\usepackage[usenames]{color}

\newtheorem{Theorem}{Theorem}
\newtheorem{Lemma}[Theorem]{Lemma}
\newtheorem{Proposition}[Theorem]{Proposition}
\newtheorem{Corollary}[Theorem]{Corollary}
\newdefinition{Definition}[Theorem]{Definition}
\newdefinition{Example}[Theorem]{Example}
\newdefinition{Remark}[Theorem]{Remark}
\newproof{proof}{Proof}
\newenvironment{Proof}{\begin{proof}}{\qed\end{proof}}

\newcommand{\pair}[2]{\{{#1},{#2}\}}
\newcommand{\two}{\mathbb{F}_2}
\newcommand{\sub}[1]{[ #1 ]}
\newcommand{\xor}{\oplus}
\newcommand{\GL}{\mathrm{GL}_2(\two)}
\renewcommand{\emptyset}{\varnothing}
\newcommand{\dual}{\mathop{\bar{*}}}
\renewcommand{\kbldelim}{(}
\renewcommand{\kbrdelim}{)}

\begin{document}


\begin{frontmatter}

\title{The Group Structure of Pivot and Loop Complementation on Graphs and Set Systems}

\author[rb]{Robert Brijder\corref{cor}}
\cortext[cor]{Corresponding author}
\ead{robert.brijder@uhasselt.be}

\author[hjh]{Hendrik Jan Hoogeboom}

\address[rb]{Hasselt University and Transnational University of Limburg, Belgium}

\address[hjh]{Leiden Institute of Advanced Computer Science,\\
Leiden University, The Netherlands}

%
%

\begin{abstract}
We study the interplay between principal pivot transform (pivot) and
loop complementation for graphs. This is done by generalizing loop
complementation (in addition to pivot) to set systems. We show that
the operations together, when restricted to single vertices, form
the permutation group $S_3$. This leads, e.g., to a normal form for
sequences of pivots and loop complementation on graphs. The results
have consequences for the operations of local complementation and
edge complementation on simple graphs: an alternative proof of a
classic result involving local and edge complementation is obtained,
and the effect of sequences of local complementations on simple
graphs is characterized.
\end{abstract}


\end{frontmatter}

\section{Introduction}
Principal pivot transform (PPT, or simply pivot), due to
Tucker~\cite{tucker1960}, partially inverts a given matrix. Its
definition is originally motivated by the extensively studied linear
complementarity problem \cite{cottle1992}. However, there are many
other application areas for PPT, see \cite{Tsatsomeros2000151} for
an overview. We consider pivots on graphs where loops are allowed
(i.e., symmetric matrices over $\two$). It is shown by
Bouchet~\cite{bouchet1987} that, in this case, the pivot operation
satisfies an equivalent definition in terms of set systems (more
specifically, in terms of delta-matroids due to a specific exchange
axiom that they fulfill).

Pivot operations on graphs (where loops are allowed) can be
decomposed into two types of elementary pivots: local
complementation and edge complementation. The names ``local
complementation'' and ``edge complementation'' are due to similar
operations on simple graphs. Local complementation on simple graphs
has originally been considered in \cite{kotzig1968} and edge
complementation has subsequently been defined in terms of local
complementation in \cite{bouchet1988}. There these operations were
motivated by circle graphs (or overlap graphs), where local and edge
complementation model natural transformations on the underlying
interval segments (or, equivalently, on Euler tours within a
4-regular graph). Many other application areas have since been
identified. For example, local complementation on simple graphs
retains the entanglement of the corresponding graph states in
quantum computing \cite{PhysRevA.69.022316}, and this operation is
of main interest in relation to rank-width in the vertex-minor
project initiated in \cite{DBLP:journals/jct/Oum05}. Moreover, edge
complementation is fundamentally related to the interlace polynomial
\cite{Arratia_InterlaceP_SODA,Arratia2004199,Aigner200411}, the
definition of which is motivated by the computation of the number of
$k$-component circuit partitions in a graph. Elementary pivots on
graphs naturally appear in the formal study of gene assembly in
ciliates~\cite{GeneAssemblyBook,BHH/PivotsDetPM/09} (a research area
of computational biology).

Surprisingly, the similarity between local and edge complementation
for simple graphs on the one hand and pivots on matrices (or graphs)
on the other hand has been largely unnoticed (although it is
observed in \cite{Geelen97}), and as a result they have been studied
almost independently.

In this paper we consider the interplay between pivots and loop
complementation (flipping the existence of loops for a given
set of vertices) on graphs. By generalizing loop
complementation to set systems, we obtain a common viewpoint
for the two operations: pivots and loop complementations are
elements of order $2$ (i.e., involutions) in the permutation
group $S_3$ (by restricting to single vertices). We find that
the dual pivot from \cite{MaxPivotsGraphs/Brijder09}
corresponds to the third element of order $2$ in $S_3$. We
obtain a normal form for sequences of pivots and loop
complementations on graphs. As a consequence a number of
results for local and edge complementations on simple graphs
are obtained including an alternative proof of a classic result
\cite{bouchet1988} relating local and edge complementation (see
Proposition~\ref{prop:class_uvu_vuv}). Finally we characterize
the effect of sequences of local complementations on simple
graphs. In this way we find that, surprisingly, loops are the
key to fully understand local and edge complementation on
simple (i.e., loopless) graphs, as they bridge the gap in the
definitions of local and edge complementation for graphs on the
one hand and simple graphs on the other.

%
An extended abstract of this paper containing selected results
without proofs was presented at TAMC~2010
\cite{BH/PivotLoopCompl/10/TAMC}.

\section{Notation and Terminology}
In this paper matrix computations (except for the first part of
Section~\ref{sec:def_pivots}) will be over $\two$, the field
consisting of two elements. We will often consider this field as the
Booleans, and its operations addition and multiplication are as such
equal to the logical exclusive-or and logical conjunction, which are
denoted by $\xor$ and $\land$ respectively. These operations carry
over to sets, e.g., for sets $A, B \subseteq V$ and $x \in V$, $x
\in A \xor B$ iff $(x \in A) \xor (x \in B)$.

A \emph{set system} (over $V$) is an ordered pair $M = (V,D)$
with $V$ a finite set and $D$ a family of subsets of $V$. We
write simply $Y \in M$ to denote $Y \in D$. For $X \subseteq
V$, $X$ is \emph{minimal} (\emph{maximal}, resp.) in $D$
w.r.t.~inclusion iff both $X \in D$ and $Y \not\in D$ for every
$Y \subset X$ ($Y \supset X$, resp.). The set of minimal
(maximal, resp.) elements of $D$ (w.r.t.~inclusion) is denoted
by $\min(D)$ ($\max(D)$, resp.). Moreover, we write $\min(M) =
\min(D)$ and $\max(M) = \max(D)$.

For a $V\times V$-matrix $A$ (the columns and rows of $A$ are
indexed by finite set $V$) and $X \subseteq V$, $A[X]$ denotes the principal
submatrix of $A$ w.r.t.~$X$, i.e., the $X \times X$-matrix obtained
from $A$ by restricting to rows and columns in $X$.

We consider undirected graphs without parallel edges, however we do
allow loops.
For graph $G=(V,E)$ we use $V(G)$ and $E(G)$ to denote its set of
vertices $V$ and set of edges $E$, respectively, where for $x \in
V$, $\{x\} \in E$ iff $x$ has a loop. For $X \subseteq V$, we denote
the subgraph of $G$ induced by $X$ as $G\sub{X}$.

With a graph $G$ one associates its adjacency matrix $A(G)$, which
is a $V \times V$-matrix $\left(a_{u,v}\right)$ over $\two$ with
$a_{u,v} = 1$ iff $\{u,v\} \in E$ (we have $a_{u,u} = 1$ iff $\{u\}
\in E$). In this way, the family of graphs with vertex set $V$
corresponds precisely to the family of symmetric $V \times
V$-matrices over $\two$. Therefore we often make no distinction
between a graph and its matrix, so, e.g., by the determinant of
graph $G$, denoted $\det G$, we will mean the determinant $\det
A(G)$ of its adjacency matrix (computed over $\two$). By convention,
$\det (G\sub{\varnothing}) = 1$.

For graph $G$, the \emph{loop complementation} operation on a set of
vertices $X \subseteq V$, denoted by $G+X$, removes loops from the vertices of $X$
when present in $G$ and adds loops to vertices of $X$ when not
present in $G$. Hence the adjacency matrix of $G+X$ is obtained from
$A(G)$ by adding $1$ to each
diagonal element $a_{xx}$, $x \in X$, of $A(G)$. Clearly,
$(G+X)+Y = G+(X\xor Y)$ for $X,Y \subseteq V$.

\section{Pivots} \label{sec:def_pivots}

In general the pivot operation is defined for matrices over
arbitrary fields, e.g., as done in \cite{Tsatsomeros2000151}. In
this paper we restrict to symmetric matrices over $\two$, which
leads to a number of additional viewpoints
to the same operation, and for each of them an
equivalent definition of the pivot operation.

\paragraph{Matrices}
Let $A$ be a $V \times V$-matrix (over an arbitrary field), and let
$X \subseteq V$ be such that the corresponding principal submatrix
$A\sub{X}$ is nonsingular, i.e., $\det A\sub{X} \neq 0$. The
\emph{pivot} of $A$ on $X$, denoted by $A*X$, is defined as follows.
If $P = A\sub{X}$ and $A =
\begin{pmatrix}
P & Q \\
R & S
\end{pmatrix}
$, then
$$
A*X = \begin{pmatrix}
P^{-1} & -P^{-1} Q \\
R P^{-1} & S - R P^{-1} Q
\end{pmatrix}.
$$
The pivot can be considered a partial inverse, as $A$ and $A*X$
satisfy the following characteristic relation, where the vectors
$x_1$ and $y_1$ correspond to the elements of $X$.
\begin{eqnarray} \label{pivot_def_reverse}
A   \begin{pmatrix} x_1 \\ x_2 \end{pmatrix} = \begin{pmatrix} y_1 \\ y_2 \end{pmatrix}
\mbox{ iff }
A*X \begin{pmatrix} y_1 \\ x_2 \end{pmatrix} = \begin{pmatrix} x_1 \\ y_2 \end{pmatrix}
\end{eqnarray}
Equality~(\ref{pivot_def_reverse}) can be used to define $A*X$ given
$A$ and $X$: any matrix $B$ satisfying this equality is of the form
$B = A*X$, see \cite[Theorem~3.1]{Tsatsomeros2000151}, and therefore
such a $B$ exists precisely when $\det A\sub{X} \neq 0$. Note that if
$\det A \not= 0$, then $A*V = A^{-1}$. Also note that by
Equation~(\ref{pivot_def_reverse}) a pivot operation is an
involution (operation of order $2$), and more generally, if
$(A*X)*Y$ is defined, then $A*(X \xor Y)$ is defined and they are
equal.

It is easy to verify that $A*X$ is skew-symmetric whenever $A$ is.
In particular, computed over $\two$, if $A$ is a graph (i.e., a
symmetric matrix over $\two$), then $A*X$ is also a graph.

The following fundamental result on pivots is due to
Tucker~\cite{tucker1960} (see also \cite{ParsonsTDP70} or
\cite[Theorem~4.1.1]{cottle1992} for an elegant proof using
Equality~(\ref{pivot_def_reverse})).

\begin{Proposition}[\cite{tucker1960}]\label{prop:tucker}
Let $A$ be a $V \times V$-matrix, and let $X\subseteq V$ be such
that $\det A\sub{X} \neq 0$. Then, for $Y \subseteq V$, $\det
(A*X)\sub{Y} = \det A\sub{X \xor Y} / \det A\sub{X}$.
\end{Proposition}

In particular, assuming that $A*X$ is defined, $(A*X)\sub{Y}$ is
nonsingular iff $A\sub{X \xor Y}$ is nonsingular.

\paragraph{Set Systems}
Let $M$ be a set system over $V$. We define, for $X \subseteq V$,
the \emph{pivot} (often called \emph{twist} in the literature, see,
e.g., \cite{Geelen97}) $M * X = (V,D * X)$, where $D * X = \{Y \xor
X \mid Y \in D\}$.

For $V \times V$-matrix $A$, let $\mathcal{M}_A = (V,D_A)$ be the
set system with $D_A = \{ X \subseteq V \mid \det A\sub{X} \neq
0\}$. As observed in \cite{bouchet1987} we have, by
Proposition~\ref{prop:tucker}, $Z \in
\mathcal{M}_{A*X}$ iff $\det ((A*X)\sub{Z}) \neq 0$ iff $\det (A\sub{X \xor
Z}) \neq 0$ iff $X \xor Z \in \mathcal{M}_A$ iff $Z \in \mathcal{M}_A * X$. Hence
$\mathcal{M}_{A*X} = \mathcal{M}_A * X$.

From now on we restrict to graphs $G$ and we work over $\two$.
Given set system $\mathcal{M}_G = (V(G),D_G)$, one can
(re)construct the graph $G$: $\{u\}$ is a loop in $G$ iff
$\{u\} \in D_G$, and $\{u,v\}$ is an edge in $G$ iff $(\{u,v\}
\in D_G) \xor ((\{u\} \in D_G) \wedge (\{v\} \in D_G))$, see
\cite[Property~3.1]{Bouchet_1991_67}. Hence the function
$\mathcal{M}_{(\cdot)}$ which assigns to each graph $G$ its set
system $\mathcal{M}_G$ is injective. In this way, the family of
graphs (with set $V$ of vertices) can be considered as a subset
of the family of set systems (over set $V$).

\begin{Remark}
Note that $\mathcal{M}_{(\cdot)}$ is not injective for binary
matrices (i.e., matrices over $\two$) in general: e.g., for
fixed $V$ with $|V| = 2$, the $2 \times 2$ zero matrix and the
matrix
$\begin{pmatrix} 0 & 1 \\
0 & 0 \end{pmatrix}$ correspond to the same set system.
Also,
$\mathcal{M}_{(\cdot)}$ is not surjective: we have, e.g.,
$\emptyset \in \mathcal{M}_{A}$ for every matrix $A$.
Consequently, the notions of binary matrix and set system are
incomparable (i.e., one is not more general than the other)
w.r.t. $\mathcal{M}_{(\cdot)}$.
\end{Remark}

As $\mathcal{M}_{G*X} = \mathcal{M}_G * X$, the pivot operation
for graphs coincides with the pivot operation for set systems.
Therefore, pivot on set systems forms an alternative definition
of pivot on graphs. Note that while for a set system $M$ over
$V$, $M * X$ is defined for all $X \subseteq V$, for a graph
$G$, $G*X$ is defined precisely when $\det G[X] = 1$, or
equivalently, when $X \in D_G$, which in turn is equivalent to
$\emptyset \in D_{G}*X$.

It turns out that $\mathcal{M}_G$ has a special structure, that of a
\emph{delta-matroid} \cite{bouchet1987}. A delta-matroid is a set
system $M$ that satisfies the symmetric exchange axiom:
For all $X,Y \in M$ and all $x \in X \xor Y$, we have $X \xor \{x\}
\in M$ or there is a $y \in X \xor Y$ with $y \not= x$ such that $X
\xor \{x,y\} \in M$\footnote{The explicit formulation of the case $X
\xor \{x\} \in M$ is often omitted in the definition of
delta-matroids. It is then understood that $y$ may be equal to $x$
and $\{x,x\} = \{x\}$. To avoid confusion we will not use this
convention here.}.
In this paper we will not use this property. In fact, we will
consider an operation on set systems that does not retain this
property of delta-matroids, cf. Example~\ref{ex:loop_comp_not_dm}.

\paragraph{Graphs}
The pivots $G*X$ where $X \in \min(D_G
\setminus \{\emptyset\})$ are called
\emph{elementary}. It is noted by Geelen~\cite{Geelen97} that an
elementary pivot $X$ corresponds to either a loop, $X = \{u\} \in
E(G)$, or to an edge, $X = \{u,v\} \in E(G)$, where (distinct)
vertices $u$ and $v$ are both non-loops. Thus for $Y \in
\mathcal{M}_G$, if $G[Y]$ has elementary pivot $X_1$, then $Y
\setminus X_1 = Y \xor X_1 \in \mathcal{M}_{G*X_1}$. By iterating
this argument, each $Y \in \mathcal{M}_G$ can be partitioned $Y =
X_1 \cup \cdots\cup X_n$ such that $G*Y = G*(X_1 \xor \cdots\xor
X_n) = (\cdots(G*X_1)\cdots * X_n)$ is a composition of elementary
pivots. Consequently, a direct definition of the elementary pivots
on graphs $G$ is sufficient to define the (general) pivot operation
on graphs.

The elementary pivot $G*\{u\}$ on a loop $\{u\}$ is called
\emph{local complementation}. It is the graph obtained from $G$ by
``toggling'' the edges in the neighbourhood $N_G(u) = \{ v \in V
\mid \{u,v\} \in E(G), u \not= v \}$ of $u$ in $G$: for each $v,w
\in N_G(u)$, $\{v,w\}\in E(G)$ iff $\{v,w\} \not\in E(G*\{u\})$, and
$\{v\}\in E(G)$ iff $\{v\} \not\in E(G*\{u\})$ (the case $v=w$). The
other edges are left unchanged.

\begin{figure}[tb]
\centerline{\unitlength 1.0mm
\begin{picture}(55,42)(0,1)
\drawccurve(02,28)(25,21)(48,32)(25,39)
\drawccurve(00,10)(10,00)(20,10)(10,20)
\drawccurve(30,10)(40,00)(50,10)(40,20)
\gasset{AHnb=0,Nw=1.5,Nh=1.5,Nframe=n,Nfill=y}
\gasset{ExtNL=y,NLdist=1.5,NLangle=90}
\put(10,02){\makebox(0,0)[cc]{$V_1$}}
\put(40,02){\makebox(0,0)[cc]{$V_2$}}
\put(25,36){\makebox(0,0)[cc]{$V_3$}}
  \node(u)(09,28){$u$}
  \node(v)(20,30){$v$}
  \node(uu)(29,32){}
  \node(vv)(41,28){}
  \node(u1)(7,14){}
  \node(u2)(14,7){}
  \node(v1)(38,7){}
  \node(v2)(43,14){}
  \drawedge(u,v){}
  \drawedge(u,u1){}
  \drawedge(u,u2){}
  \drawedge(v,v1){}
  \drawedge(v,v2){}
  \drawedge(u1,v2){}
  \drawedge(u2,v1){}
  \drawedge[dash={1}0](v1,v2){}
  \drawedge[dash={1}0](u1,u2){}
  \drawedge[dash={1}0](uu,vv){}
  \drawedge(uu,u1){}
  \drawedge(vv,u2){}
  \drawedge(uu,v1){}
  \drawedge(vv,v2){}
\end{picture}
\begin{picture}(55,42)(0,1)
\drawccurve(02,28)(25,21)(48,32)(25,39)
\drawccurve(00,10)(10,00)(20,10)(10,20)
\drawccurve(30,10)(40,00)(50,10)(40,20)
\gasset{AHnb=0,Nw=1.5,Nh=1.5,Nframe=n,Nfill=y}
\gasset{ExtNL=y,NLdist=1.5,NLangle=90}
\put(10,02){\makebox(0,0)[cc]{$V_1$}}
\put(40,02){\makebox(0,0)[cc]{$V_2$}}
\put(25,36){\makebox(0,0)[cc]{$V_3$}}
  \node(u)(09,28){$u$}
  \node(v)(20,30){$v$}
  \node(uu)(29,32){}
  \node(vv)(41,28){}
  \node(u1)(7,14){}
  \node(u2)(14,7){}
  \node(v1)(38,7){}
  \node(v2)(43,14){}
  \drawedge(u,v){}
  \drawedge(v,u1){}
  \drawedge(v,u2){}
  \drawedge(u,v1){}
  \drawedge(u,v2){}
  \drawedge(u1,v1){}
  \drawedge(u2,v2){}
  \drawedge[dash={1}0](v1,v2){}
  \drawedge[dash={1}0](u1,u2){}
  \drawedge[dash={1}0](uu,vv){}
  \drawedge(uu,u2){}
  \drawedge(vv,u1){}
  \drawedge(uu,v2){}
  \drawedge(vv,v1){}
\end{picture}%
}
\caption{Pivot on an edge $\pair uv$ in a graph.
 Adjacency between vertices $x$ and $y$ is toggled iff $x\in V_i$ and $y\in V_j$ with
 $i\neq j$. Note that $u$ and $v$ are adjacent to all vertices in $V_3$ ---
 these edges are omitted in the diagram. The operation does not
 affect edges adjacent to vertices outside the sets
 $V_1,V_2,V_3$, nor does it change any of the loops.}%
\label{fig:pivot}
\end{figure}
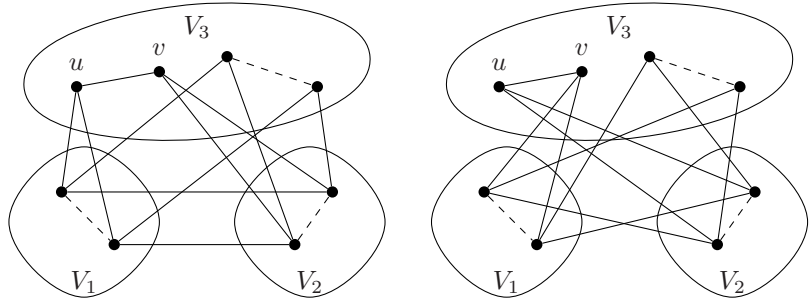

We now recall \emph{edge complementation} $G*\pair uv$ on an edge
$\pair uv$ between non-loop vertices. For a vertex $x$ consider its
closed neighbourhood $N'_G(x)= N_G(x)\cup \{x\}$. The edge $\pair
uv$ partitions the vertices of $G$ adjacent to $u$ or $v$ into
three sets $V_1 = N'_G(u) \setminus N'_G(v)$, $V_2 = N'_G(v)
\setminus N'_G(u)$, $V_3 = N'_G(u) \cap N'_G(v)$. Note that $u,v \in
V_3$.

The graph  $G*\pair uv$ is constructed by ``toggling'' all edges
between different $V_i$ and $V_j$: for $\pair xy$ with $x\in V_i$
and $y\in V_j$ ($i\neq j$): $\pair xy \in E(G)$ iff $\pair xy \notin
E(G*{\{u,v\}})$, see Figure~\ref{fig:pivot}. The other edges
remain unchanged. Note that, as a result of this operation, the
neighbours of $u$ and $v$ are interchanged.

\newcommand{\pivotorbitmacro}{
  \gasset{AHnb=0,Nw=1.5,Nh=1.5,Nframe=n,Nfill=y}
  \gasset{AHnb=0,Nw=8,Nh=8,Nframe=y,Nfill=n}
    \node(r)(05,05){$r$}
    \node(p)(05,25){$p$}
    \node(q)(25,25){$q$}
    \node(s)(25,05){$s$}
}

\begin{figure}[t]
\unitlength 1mm%
\begin{center}
\resizebox{\textwidth}{!}{
\begin{picture}(135,53)(-27,-8)
\node[Nw=25,Nh=25,Nframe=n](I)(00,35){%
  \unitlength0.4mm
  \begin{picture}(30,30)
  \pivotorbitmacro
  \drawedge(p,q){}
  \drawedge(p,r){}
  \drawedge(p,s){}
  \drawedge(r,s){}
  \drawloop[loopangle=45,loopdiam=7](q){}
  \drawloop[loopangle=135,loopdiam=7](p){}
  \end{picture}
}
\node[Nw=25,Nh=25,Nframe=n](II)(80,35){%
  \unitlength0.4mm
  \begin{picture}(30,30)
  \pivotorbitmacro
  \drawedge(p,q){}
  \drawedge(p,r){}
  \drawedge(p,s){}
  \drawedge(r,s){}
  \drawloop[loopangle=45,loopdiam=7](q){}
\end{picture}
}
\node[Nw=25,Nh=25,Nframe=n](III)(80,0){%
  \unitlength0.4mm
  \begin{picture}(30,30)
  \pivotorbitmacro
  \drawedge(p,r){}
  \drawedge(p,s){}
  \drawedge(q,r){}
  \drawedge(q,s){}
  \drawedge(r,s){}
  \drawloop[loopangle=45,loopdiam=7](q){}
\end{picture}
}
\node[Nw=25,Nh=25,Nframe=n](IV)(40,00){%
  \unitlength0.4mm
  \begin{picture}(30,30)
  \pivotorbitmacro
  \drawedge(p,r){}
  \drawedge(p,s){}
  \drawedge(q,r){}
  \drawedge(q,s){}
  \drawloop[loopangle=45,loopdiam=7](q){}
  \drawloop[loopangle=225,loopdiam=7](r){}
  \drawloop[loopangle=-45,loopdiam=7](s){}
\end{picture}
}
\node[Nw=25,Nh=25,Nframe=n](V)(00,00){%
  \unitlength0.4mm
  \begin{picture}(30,30)
  \pivotorbitmacro
  \drawedge(p,q){}
  \drawedge(p,r){}
  \drawedge(p,s){}
  \drawedge(q,r){}
  \drawedge(q,s){}
  \drawloop[loopangle=225,loopdiam=7](r){}
  \drawloop[loopangle=135,loopdiam=7](p){}
  \drawloop[loopangle=-45,loopdiam=7](s){}
\end{picture}
}
  \drawedge[AHnb=1,ATnb=1](I,II){$*\{q\}$}
  \drawloop[loopangle=180,loopdiam=8](I){$*\{r,s\}$}
  \drawloop[loopangle=0,loopdiam=8](II){$*\{r,s\}$}
  \drawedge[AHnb=1,ATnb=1,sxo=4,exo=4](II,III){$*\{p,r\}$}
  \drawedge[AHnb=1,ATnb=1,sxo=-4,exo=-4,ELside=r](II,III){$*\{p,s\}$}
  \drawloop[loopangle=00,loopdiam=8](III){$*\{r,s\}$}
  \drawedge[AHnb=1,ATnb=1](III,IV){$*\{q\}$}
  \drawedge[AHnb=1,ATnb=1,syo=4,eyo=4](IV,V){$*\{r\}$}
  \drawedge[AHnb=1,ATnb=1,syo=-4,eyo=-4](IV,V){$*\{s\}$}
  \drawedge[AHnb=1,ATnb=1](V,I){$*\{p\}$}
\end{picture}
}
\end{center}
\caption{The orbit of $G$ of Example~\ref{ex:introd_pivot} under
pivot. Only the elementary pivots are shown.}\label{fig:pivot_space}
\end{figure}
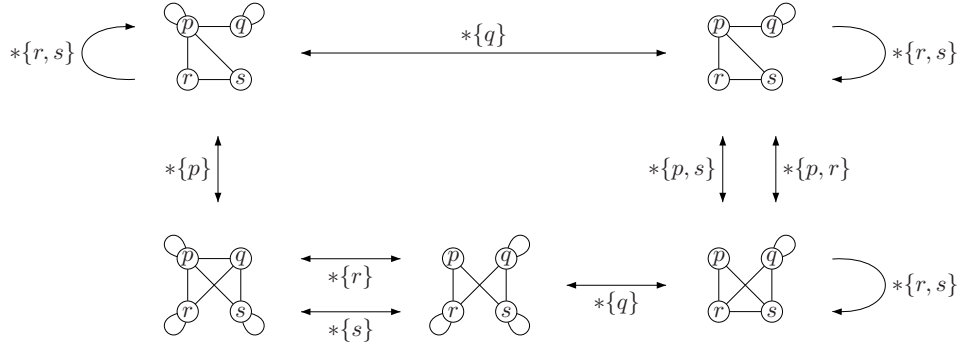

\begin{Example}\label{ex:introd_pivot}
Let $G$ be the graph depicted in the upper-left corner of
Figure~\ref{fig:pivot_space}.
We have $A(G) = \kbordermatrix{ & p & q & r & s \\
p & 1 & 1 & 1 & 1 \\
q & 1 & 1 & 0 & 0 \\
r & 1 & 0 & 0 & 1 \\
s & 1 & 0 & 1 & 0 }$.
Graph $G$ corresponds to $\mathcal{M}_G = (\{p,q,r,s\},D_G)$, where
$$
D_G = \{\varnothing, \{p\}, \{q\}, \{p,r\}, \{p,s\}, \{r,s\},
\{p,q,r\}, \{p,q,s\}, \{p,r,s\}, \{q,r,s\} \}.
$$
For example, $\{p,r\} \in D_G$ since $\det(G[\{p,r\}]) =
\det \begin{pmatrix}
1 & 1 \\
1 & 0
\end{pmatrix} =
1$. The orbit of $G$ under pivot as well as the applicable
elementary pivots (i.e., local and edge complementation) are shown
in Figure~\ref{fig:pivot_space}. For example, $G * \{p,q,r\}$ is
shown on the lower-right in the same figure. Note that $D_G *
\{p,q,r\} =$
$$
\{ \varnothing, \{q\}, \{p,r\}, \{p,s\}, \{q,r\},
\{q,s\}, \{r,s\}, \{p,q,r\}, \{p,q,s\}, \{q,r,s\} \}
$$
indeed corresponds to $G
* \{p,q,r\}$.
\end{Example}

\section{Unifying Pivot and Loop Complementation}
\label{sec:unifying}

We now introduce a class of operations on set systems. \ifshort It
\else As we will show, it \fi turns out that this class contains
both the pivot and (a generalization of) loop complementation. Each
operation is a linear transformation, where the input and output
vectors indicate the presence (or absence) of sets $Z$ and $Z\setminus\{j\}$
in the original and resulting set systems.

\begin{Definition} \label{def:vertex_flip}
Let $M = (V,D)$ be a set system, and let $\alpha$ be a $2 \times
2$-matrix over $\two$. We define, for $j \in V$, the \emph{vertex
flip} $\alpha$ of $M$ on $j$, denoted by $M\alpha^j = (V,D')$,
where, for all $Z \subseteq V$ with $j \in Z$, the membership of $Z$
and $Z\setminus\{j\}$ in $D'$ is determined as follows:
$$
\alpha \; (Z \in D, Z\setminus\{j\} \in D )^T = (Z \in D', Z\setminus\{j\} \in D' )^T.
$$
\end{Definition}

\newcommand{\melu}{a_{11}}
\newcommand{\meld}{a_{21}}
\newcommand{\meru}{a_{12}}
\newcommand{\merd}{a_{22}}
\newcommand{\metlu}{b_{11}}
\newcommand{\metld}{b_{21}}
\newcommand{\metru}{b_{12}}
\newcommand{\metrd}{b_{22}}

In the above definition, we regard the elements of the vectors as
Boolean values, e.g., the expression $Z \in D$ obtains
either true ($1$) or false
($0$). To be more explicit, let $\alpha = \begin{pmatrix} \melu & \meru \\
\meld & \merd \end{pmatrix}$. Then we have for all $Z
\subseteq V$, $Z \in D'$ iff
$$\begin{cases}
(\melu \wedge Z \in D) \xor (\meru \wedge Z \setminus \{j\} \in D) & \mbox{if } j \in Z \\
(\meld \wedge Z \cup \{j\} \in D) \xor (\merd \wedge Z \in D) & \mbox{if } j \not\in Z \\
\end{cases}.$$
Note that in the above statement we may replace both $Z \cup \{j\}
\in D$ and $Z \setminus \{j\} \in D$ by $Z \xor \{j\} \in D$ as in the
former we have $j \not\in Z$ and in the latter we have $j \in Z$.
Thus, the operation $\alpha^j$ decides whether or not set $Z$ is in
the new set system, based on the fact whether or not $Z$ and $Z \xor
\{j\}$ belong to the original system.

Note that if $\alpha$ is the identity matrix, then $\alpha^j$ is
simply the identity operation. Moreover, with $\alpha_{*} = \begin{pmatrix} 0 & 1
\\ 1 & 0
\end{pmatrix}$
we have $M\alpha^j_{*} = M * \{j\}$, the pivot operation on a single
element $j$.

By definition, a composition of vertex flips on the same element
corresponds to matrix multiplication. \ifshort The \else Moreover,
the \fi following lemma shows that vertex flips on different
elements commute.
\begin{Lemma}
\label{lem:matrix_matroid_ops} Let $M$ be a set system over $V$, and
let $j,k \in V$. We have that $(M\alpha^j)\beta^j =
M(\beta\alpha)^j$, where $\beta\alpha$ denotes matrix multiplication
of $\beta$ and $\alpha$. Moreover $(M\alpha^j)\beta^k =
(M\beta^k)\alpha^j$ if $j \neq k$.
\end{Lemma}
\ifshort \else
\begin{Proof}
The fact that $(M\alpha^j)\beta^j = M(\beta\alpha)^j$ follows
directly from Definition~\ref{def:vertex_flip}.

Let $M = (V,D)$, and assume that $j \neq k$. Let $M\alpha^j = (V,D')$,
and let $M\beta^k = (V,D'')$. For any set $Z \subseteq V$ with $j,k
\in Z$, we consider the sets $Z$, $Z\setminus\{j\}$, $Z\setminus\{k\}$, and
$Z\setminus\{j,k\}$. Now, for any family $Q$ of subsets of $V$, let
$v_Q = (Z \in Q, Z\setminus\{j\} \in Q, Z\setminus\{k\} \in Q,
Z\setminus\{j,k\} \in Q)^T$. The $4 \times 4$-matrices $\alpha'$ and
$\beta'$ such that $\alpha' v_D = v_{D'}$ and $\beta' v_D = v_{D'}$,
are

$\alpha' = \left( \begin{array}{*4c}
\melu & \meru & 0      & 0\\
\meld & \merd & 0      & 0\\
0      & 0      & \melu & \meru \\
0      & 0      & \meld & \merd
\end{array} \right)$
and
$\beta' = \left( \begin{array}{*4c}
\metlu & 0      & \metru & 0\\
0      & \metlu & 0      & \metru\\
\metld & 0      & \metrd & 0\\
0      & \metld & 0      & \metrd
\end{array} \right)$,

\noindent where

$\alpha= \left( \begin{array}{cc} \melu & \meru\\
\meld & \merd
\end{array} \right)$
and
$\beta= \left( \begin{array}{cc} \metlu & \metru\\
\metld & \metrd
\end{array} \right)$.

Equivalently, focussing on the $2\times 2$ blocks,
we have
$\alpha' = \left(
\begin{array}{cc} \alpha & 0 \\
0 & \alpha \end{array}
\right)$
and
$\beta' = \left(
\begin{array}{cc} \metlu I  & \metru I\\
 \metld I  & \metrd I  \end{array}
\right)$.
It is easy to see these matrices commute. Multiplication (in either
order) yields the $4\times4$-matrix $\alpha' \beta' = \beta' \alpha'
= \left(
\begin{array}{cc} (\metlu I) \alpha  & (\metru I) \alpha\\
(\metld I) \alpha  & (\metrd I) \alpha  \end{array} \right)$.
\end{Proof}
\fi
To simplify notation, we assume left associativity of the vertex
flip, and write $M \varphi_1 \varphi_2 \cdots \varphi_n$ to denote
$(\cdots((M \varphi_1) \varphi_2) \cdots ) \varphi_n$, where
$\varphi_1 \varphi_2 \cdots \varphi_n$ is a sequence of vertex flip
operations applied to set system $M$. Hence, as a special case of
the vertex flip, the pivot operation is also written in the
simplified notation. We carry this simplified notation over to
graphs $G$.

Due to the commutative property shown in
Lemma~\ref{lem:matrix_matroid_ops} we (may) define, for a set $X =
\{x_1, \dots , x_n\} \subseteq V$, $M \alpha^X = M \alpha^{x_1}
\alpha^{x_2} \cdots \alpha^{x_n}$, where the result is independent
of the order in which the operations are applied. Moreover, if
$\alpha$ is of order $2$ (i.e., $\alpha \alpha$ is the identity
matrix), then $(M \alpha^X)\alpha^Y = M \alpha^{X \xor Y}$.

Now consider
$\alpha_+ = \left( \begin{array}{cc} 1 & 1 \\
0 & 1 \end{array} \right)$. The matrices $\alpha_+$ and $\alpha_{*}$
given above generate the group $\GL$ of $2\times2$ matrices with
non-zero determinant. In fact $\GL$ is isomorphic to the group $S_3
= \{1, a, b, c, f, g\}$ of permutations of three elements, where $1$
is the identity, $a$, $b$, and $c$ are the elements of order $2$,
and $f$ and $g$ are the elements of order $3$. The matrices
$\alpha_+$ and $\alpha_{*}$ are both of order $2$ and we may
identify them with any two (distinct) elements of $S_3$ of order
$2$. The generators \ifshort\else $\alpha_+$ and $\alpha_{*}$ \fi
satisfy the relations $\alpha_+^2=1$,  $\alpha_*^2=1$,  and
$(\alpha_*\alpha_+)^3=1$.

As, by Lemma~\ref{lem:matrix_matroid_ops}, vertex flips on $j$ and
$k$ with $j \not= k$ commute, we have that the vertex flips form the
group $(S_3)^V$ of functions $f: V \rightarrow S_3$ where
composition/multiplication is point wise: $(f g)(j) = f(j)g(j)$ for
all $j \in V$. Note that by fixing a linear order of $V$, $(S_3)^V$
is isomorphic to $(S_3)^n$ with $n = |V|$, the direct product of $n$
times group $S_3$. The vertex flips form an action of $(S_3)^V$ on
the family of set systems over $V$.

\section{Loop Complementation and Set Systems}
In this section we focus on vertex flips of matrix $\alpha_+$
(defined in the previous section). We \ifshort\else will~\fi show
that this operation is a generalization to set systems of loop
complementation for graphs (cf.
Theorem~\ref{thm:matroid_analog_loop_complementation}). Consequently,
we will call it \emph{loop complementation} as well.

Let $M = (V,D)$ be a set system and $j \in V$. We denote
$M\alpha^j_+$ by $M+\{j\}$. Hence, we have $M+\{j\} = (V,D')$ where,
for all $Z \subseteq V$, $Z \in D'$ iff \ifshort 1) $(Z \in D) \xor
(Z \setminus \{j\} \in D)$ if $j \in Z$, and 2) $Z \in D$ if $j \not\in Z$.
\else
$$\begin{cases}
(Z \in D) \xor (Z \setminus \{j\} \in D) & \mbox{if } j \in Z \\
Z \in D & \mbox{if } j \not\in Z \\
\end{cases}.$$\fi

The definition of loop complementation can be reformulated as
follows:
$D' = D \xor \{X \cup \{j\} \mid X \in D, j \not\in
X\}$.

\begin{Example} \label{ex:loop_compl_set_system}
Let $V = \{1,2,3\}$ and $M = (V,\{\varnothing, \{1\}, \{1,2\},
\{3\}, \{1,2,3\} \})$ be a set system. We have $M$ $+$ $\{3\}$ $=$
$(V,\{\varnothing, \{1\},
\{1,2\}, \{3\},$ $\{1,2,3\} \}$ $\xor$ $\{ \{3\}, \{1,3\},$ $\{1,2,3\}
\})$ $=$ $(V,$ $\{\varnothing, \{1\},$ $\{1,2\},$ $\{1,3\} \})$.
\end{Example}

We denote, for $X \subseteq V$, $M\alpha^X_+$ by $M+X$. Moreover, as
$\alpha_+$ is of order $2$, we have, similar to the pivot operation,
$(M+X)+Y = M+(X \xor Y)$. Also, by the commutative property of
vertex flip in Lemma~\ref{lem:matrix_matroid_ops}, we have for $X,Y
\subseteq V$ with $X \cap Y = \emptyset$, $M*X+Y = M+Y*X$.

We now provide a characterization of loop complementation which describes how successive applications of loop complementation in set systems interact.

\begin{Theorem} \label{thm:loopc_ss_char}
Let $M$ be a set system and $X,Y \subseteq V$. We have $Y \in M+X$ iff $|\{ Z \in M \mid Y \setminus X \subseteq Z \subseteq Y \}|$ is odd.
\end{Theorem}
\begin{Proof}
The proof is by induction on $|X|$. First consider the case $X = \emptyset$.
$Y \in M+\emptyset$  iff
$Y \in M$ iff
$|\{ Z \in M \mid Z=Y \}|$ is odd.

Now consider $X \cup \{y\}$ with $y\notin X$ in the induction step.

If $y \notin Y$, then $Y \in M+X+\{y\}$
iff
$Y \in M+X$
iff
$|\{ Z \in M \mid Y\setminus X \subseteq Z \subseteq Y \}|$ is odd
iff
$|\{ Z \in M \mid (Y\setminus X)\setminus\{y\} \subseteq Z \subseteq Y \}|$ is odd,
as $Y\setminus X = (Y\setminus X)\setminus\{y\}$.

Now assume that $y \in Y$.
Let
$C_1 = \{ Z \in M \mid (Y\setminus\{y\})\setminus X \subseteq Z \subseteq Y\setminus\{y\} \}$
and let
$C_2 = \{ Z \in M \mid Y\setminus X \subseteq Z \subseteq Y\}$.
Elements in $C_1$ do not contain $y$ whereas those in $C_2$ do.
Thus $C_1$ and $C_2$ are disjoint, and
$C_1 \cup C_2 = \{ Z \in M \mid Y\setminus(X\cup\{y\}) \subseteq Z \subseteq Y \}$.
Moreover $|C_1\cup C_2|$ is odd iff exactly one of
$|C_1|$ and $|C_2|$ is odd.

By definition of loop complementation
$Y \in (M+X)+y$ iff
($Y\setminus\{y\} \in M+X) \xor (Y \in M+X)$.
According to the induction hypothesis this means that exactly one of
$|C_1|$ and $|C_2|$ is odd, i.e.,
$|\{ Z \in M \mid Y\setminus(X\cup\{y\}) \subseteq Z \subseteq Y \}|$ is odd,
as required.
\end{Proof}

The next result implies that indeed the notion of loop
complementation for set systems is a generalization of the
notion of loop complementation for graphs.

\begin{Theorem}
\label{thm:matroid_analog_loop_complementation} Let $A$ be a $V
\times V$-matrix over $\two$ and $X \subseteq V$. Then
$\mathcal{M}_{A+X} = \mathcal{M}_{A}+X$.
\end{Theorem}
\ifshort\else
\begin{Proof}
It suffices to show the result for $X = \{j\}$ with $j \in V$, as
the general case follows by the commutative property of vertex flip
(Lemma~\ref{lem:matrix_matroid_ops}). Let $Z\subseteq V$. We compare
$\det A[Z]$ with $\det (A+\{j\})[Z]$. First assume that $j\notin Z$. Then
$A[Z] = (A+\{j\})[Z]$, thus $\det A[Z] = \det (A+\{j\}) [Z]$. Now
assume that $j\in Z$, which implies that $A[Z]$ and $(A+\{j\})[Z]$ differ
in exactly one position: $(j,j)$. We may compute determinants by
Laplace expansion over the $j$-th column, and summing minors. As
$A[Z]$ and $(A+\{j\})[Z]$ differ at only the matrix-element $(j,j)$,
these expansions differ only in the inclusion of minor $\det
A[Z\setminus\{j\}]$. Thus $\det (A+\{j\})[Z]$ equals $\det A[Z] \xor \det
A[Z\setminus\{j\}]$, from which the statement follows.
\end{Proof}
\fi
Surprisingly, this natural definition of loop complementation
on set systems is not found in the literature.

\begin{figure}
\unitlength 0.50mm
\centerline{%
\begin{picture}(250,120)(0,35)\unitlength 0.50mm
\node[Nframe=n,Nw=55,Nh=60](EEN)(25,75){\unitlength 0.50mm
\begin{picture}(50,60)(5,0)
\gasset{AHnb=0}
\gasset{Nw=10,Nh=10,Nmr=5,Nframe=y,loopdiam=8}
\node(p)(30,95){$1$}
\node(q)(15,80){$2$}
\node(r)(45,80){$3$}
\drawedge(p,q){}
\drawedge(p,r){}
\drawloop[loopangle=90](p){}
\drawloop[loopangle=-20](r){}
\gasset{Nw=10,Nh=10,Nmr=5,Nframe=n}
\node(011)(35,30){\small$2$}
\node[Nframe=y](010)(35,00){$\emptyset$}
\node(001)(10,40){\small$23$}
\node[Nframe=y](000)(10,10){\small$3$}
\node[Nframe=y](111)(50,45){\small$12$}
\node[Nframe=y](110)(50,15){\small$1$}
\node[Nframe=y](101)(25,55){\small$123$}
\node(100)(25,25){\small$13$}
\drawedge(000,001){}\drawedge(010,011){}
\drawedge(000,010){}\drawedge(001,011){}
\drawedge(100,101){}\drawedge(110,111){}
\drawedge(100,110){}\drawedge(101,111){}
\drawedge(100,000){}\drawedge(101,001){}
\drawedge(110,010){}\drawedge(111,011){}
\end{picture}}
\node[Nframe=n,Nw=50,Nh=60](TWEE)(90,75){\unitlength 0.50mm
\begin{picture}(50,60)(5,0)
\gasset{AHnb=0}
\gasset{Nw=10,Nh=10,Nmr=5,Nframe=y,loopdiam=8}
\node(p)(30,95){$1$}
\node(q)(15,80){$2$}
\node(r)(45,80){$3$}
\drawedge(p,q){}
\drawedge(p,r){}
\drawloop[loopangle=90](p){}
\gasset{Nw=10,Nh=10,Nmr=5,Nframe=n}
\node(011)(35,30){\small$2$}
\node[Nframe=y](010)(35,00){$\emptyset$}
\node(001)(10,40){\small$23$}
\node(000)(10,10){\small$3$}
\node[Nframe=y](111)(50,45){\small$12$}
\node[Nframe=y](110)(50,15){\small$1$}
\node(101)(25,55){\small$123$}
\node[Nframe=y](100)(25,25){\small$13$}
\drawedge(000,001){}\drawedge(010,011){}
\drawedge(000,010){}\drawedge(001,011){}
\drawedge(100,101){}\drawedge(110,111){}
\drawedge(100,110){}\drawedge(101,111){}
\drawedge(100,000){}\drawedge(101,001){}
\drawedge(110,010){}\drawedge(111,011){}
\end{picture}}
\node[Nframe=n,Nw=50,Nh=60](DRIE)(155,75){\unitlength 0.50mm
\begin{picture}(50,60)(5,0)
\gasset{AHnb=0}
\gasset{Nw=10,Nh=10,Nmr=5,Nframe=y,loopdiam=8}
\node(p)(30,95){$1$}
\node(q)(15,80){$2$}
\node(r)(45,80){$3$}
\drawedge(p,q){}
\drawedge(p,r){}
\gasset{Nw=10,Nh=10,Nmr=5,Nframe=n}
\node(011)(35,30){\small$2$}
\node[Nframe=y](010)(35,00){$\emptyset$}
\node(001)(10,40){\small$23$}
\node(000)(10,10){\small$3$}
\node[Nframe=y](111)(50,45){\small$12$}
\node(110)(50,15){\small$1$}
\node(101)(25,55){\small$123$}
\node[Nframe=y](100)(25,25){\small$13$}
\drawedge(000,001){}\drawedge(010,011){}
\drawedge(000,010){}\drawedge(001,011){}
\drawedge(100,101){}\drawedge(110,111){}
\drawedge(100,110){}\drawedge(101,111){}
\drawedge(100,000){}\drawedge(101,001){}
\drawedge(110,010){}\drawedge(111,011){}
\end{picture}}
\node[Nframe=n,Nw=55,Nh=60](VIER)(220,75){\unitlength 0.50mm
\begin{picture}(50,60)(5,0)
\gasset{AHnb=0}
\gasset{Nw=10,Nh=10,Nmr=5,Nframe=y,loopdiam=8}
\node(p)(30,95){$1$}
\node(q)(15,80){$2$}
\node(r)(45,80){$3$}
\drawedge(p,q){}
\drawedge(p,r){}
\drawloop[loopangle=200](q){}
\gasset{Nw=10,Nh=10,Nmr=5,Nframe=n}
\node[Nframe=y](011)(35,30){\small$2$}
\node[Nframe=y](010)(35,00){$\emptyset$}
\node(001)(10,40){\small$23$}
\node(000)(10,10){\small$3$}
\node[Nframe=y](111)(50,45){\small$12$}
\node(110)(50,15){\small$1$}
\node[Nframe=y](101)(25,55){\small$123$}
\node[Nframe=y](100)(25,25){\small$13$}
\drawedge(000,001){}\drawedge(010,011){}
\drawedge(000,010){}\drawedge(001,011){}
\drawedge(100,101){}\drawedge(110,111){}
\drawedge(100,110){}\drawedge(101,111){}
\drawedge(100,000){}\drawedge(101,001){}
\drawedge(110,010){}\drawedge(111,011){}
\end{picture}}
\gasset{AHnb=1,ATnb=1,linewidth=1.3,linecolor=Gray,ATangle=30,ATLength=6,ATlength=3,AHangle=30,AHLength=6,AHlength=3}
\drawedge(EEN,TWEE){\large${}+3$}
\drawedge(TWEE,DRIE){\large${}+1$}
\drawedge(DRIE,VIER){\large${}+2$}
\end{picture}%
}
\caption{Toggling one-by-one loops on the vertices of a graph, and the
corresponding set systems.
}
\label{fig:loop_compl_graph}
\end{figure}
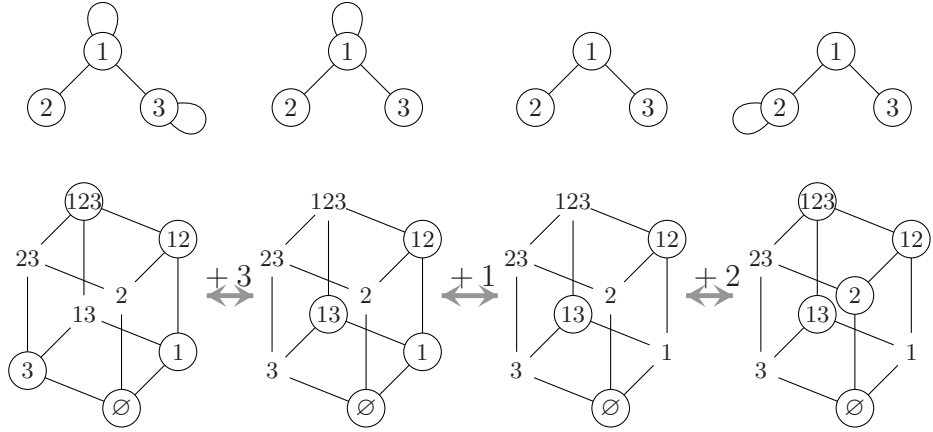

\begin{Example} \label{ex:loop_compl_graph}
The set system $M = (\{1,2,3\},\{\varnothing, \{1\}, \{1,2\}, \{3\},
\{1,2,3\} \})$ of Example~\ref{ex:loop_compl_set_system} has a graph
representation $G$: $M = \mathcal{M}_G$ and $G$ are given on the
left-hand side in Figure~\ref{fig:loop_compl_graph}. The figure
also contains some other set systems obtainable from $M$ through loop
complementation. Notice that
$M+\{3\} = (\{1,2,3\},\{\varnothing, \{1\}, \{1,2\}, \{1,3\} \})$ of
Example~\ref{ex:loop_compl_set_system}
corresponds to graph $G+\{3\}$.
\end{Example}

While for a set system the property of being a delta-matroid is closed under
pivot, the next example shows that it is \emph{not} closed under loop
complementation.
%

\begin{Example} \label{ex:loop_comp_not_dm}
Let $V = \{1,2,3\}$ and $M = (V,D)$ with $D = \{\emptyset, \{1\},
\{2\}, \{3\},$ $\{1,2\},$ $\{2,3\},$ $\{3,1\}\}$ be a set system. It
is shown in \cite[Section~3]{Bouchet_1991_67} that $M$ is a
delta-matroid without graph representation. Consider $\{1\}
\subseteq V$. Then $M+\{1\} = (V,D')$ with $D' =
\{\emptyset,\{2\},\{3\},\{2,3\},\{1,2,3\}\}$ is not a delta-matroid:
for $X = \emptyset, Y = \{1,2,3\} \in D'$, and $x = 1 \in X \xor Y$,
we have $X \xor \{x\} = \{1\} \not\in D'$ and there is no $y \in X
\xor Y$ such that $X \xor \{x,y\} \in D'$.
\end{Example}

\section{Compositions of Loop Complementation and Pivot}

In this section we study sequences of loop complementation and pivot
operations. As we may consider both operations as vertex flips, we
obtain in a straightforward way general equalities involving loop
complementation and pivot.

\begin{Theorem} \label{thm:main_result_matroids}
Let $M$ be a set system over $V$ and $X \subseteq V$. Then $M+X*X+X
= M*X+X*X$.
\end{Theorem}
\ifshort\else
\begin{Proof}
In group $S_3$ we have $a b a = b a b = c$. Hence $\alpha_+ \alpha_*
\alpha_+ = \alpha_* \alpha_+ \alpha_*$. Now by
Lemma~\ref{lem:matrix_matroid_ops}, we have $M+\{j\}*\{j\}+\{j\} =
M*\{j\}+\{j\}*\{j\}$ for any $j \in V$. By the commutative property
of vertex flip in Lemma~\ref{lem:matrix_matroid_ops}, this can be
generalized to sets $X \subseteq V$, and hence we obtain the desired
result.
\end{Proof}
\fi

Let us denote $\alpha_{\dual} = \alpha_+ \alpha_* \alpha_+$ and
denote, for $X \subseteq V$, $M\alpha^X_{\dual}$ by $M\dual X$.
We will call the $\dual$ operation the \emph{dual pivot}. As
$\alpha_+$ is of order $2$, we have, similar to the pivot operation
and loop complementation, $(M\dual X)\dual Y = M\dual (X \xor
Y)$. The dual pivot together with pivot and loop complementation
correspond precisely to the elements of order $2$ in $S_3$.

We now obtain a normal form for sequences of pivots and loop
complementations.

\begin{Theorem} \label{thm:normal_form_pivot_loopc_matroids}
Let $M$ be a set system over $V$, and let $\varphi$ be any sequence
of pivot and loop complementation operations on elements in $V$. We
have that $M \varphi = M+X*Y+Z$ for some $X, Y, Z \subseteq V$ with
$X \subseteq Y$.
\end{Theorem}
\ifshort\else
\begin{Proof}
Again we can consider the operations with respect to a single
element $j$, as the generalization to sets follows by the
commutative property of Lemma~\ref{lem:matrix_matroid_ops}.

The $6$ elements of $\GL$ are $1$, $\alpha_+$, $\alpha_*$,
$\alpha_+ \alpha_*$, $\alpha_* \alpha_+$, and $\alpha_+ \alpha_*
\alpha_+$. Hence any sequence of pivot and loop complementation over
$j$ reduces to one of these six elements, each of which can be
written in the form of the statement (with $X,Y,Z$
either equal to $\{j\}$ or to the empty set).
\end{Proof}
\fi

Because local and edge complementation operations are special cases
of pivot the normal form of
Theorem~\ref{thm:normal_form_pivot_loopc_matroids} is equally valid
for any sequence $\varphi$ of local, edge, and loop complementation
operations.

The central interest of this paper is to study compositions of pivot
and loop complementation on graphs. As explained in
Section~\ref{sec:def_pivots}, the pivot operations for set systems
and graphs coincide, i.e., $\mathcal{M}_{G*X} = \mathcal{M}_G*X$,
and we have taken care that the same holds for loop complementation,
cf. Theorem~\ref{thm:matroid_analog_loop_complementation}. Hence
results that hold for set systems in general, like
Theorem~\ref{thm:main_result_matroids}, subsume the special case
where the set system $M$ represents a graph (i.e., $M =
\mathcal{M}_G$ for some graph $G$)
--- recall that the injectivity
of $\mathcal{M}_{(\cdot)}$ allows one to view the family
$\mathcal{G}$ of graphs (over $V$) as a subset of the family of set
systems (over $V$). We only need to make sure that we ``stay'' in
$\mathcal{G}$, i.e., by applying a pivot or loop complementation
operation to $\mathcal{M}_G$ we obtain a set system $M$ such that $M
= \mathcal{M}_{G'}$ for some graph $G'$. For loop complementation
this will always hold, however care must be taken for pivot as
$\mathcal{M}_G*X$, which is defined for all $X \subseteq V$, only
represents a graph if $\det G\sub{X} = 1$. Hence when restricting a
general result (on pivot or local complementation for set systems)
to graphs, we add the condition of applicability of the operations.


It is useful to explicitly state
Theorem~\ref{thm:main_result_matroids} restricted to graphs.
This is a fundamental result for pivots on graphs (or,
equivalently, symmetric matrices over $\two$) not found in the
literature. We will study some of its consequences in the
remainder of this paper.
\begin{Corollary} \label{cor:main_result_graphs}
Let $G$ be a graph and $X \subseteq V$. Then $G+X*X+X = G*X+X*X$
when both sides are defined.
\end{Corollary}

In the particular case of Corollary~\ref{cor:main_result_graphs} it
is not necessary to verify the applicability of both sides: it turns
out that the applicability of the right-hand side implies the
applicability of the left-hand side of the equality.
\begin{Lemma}
\label{lem:def_left_right_pivots} Let $G$ be a graph and $X
\subseteq V$. If $G*X+X*X$ is defined, then $G+X*X+X$ is defined.
\end{Lemma}
\ifshort\else
\begin{Proof}
Assume that $G*X+X*X$ is defined. Thus, $G_2 = G_1+X*X+X*X+X$ is
defined for $G_1 = G+X$. Now consider $\mathcal{M}_{G_2}$. We have
that $\mathcal{M}_{G_2}*X = \mathcal{M}_{G_1}$ by
Theorem~\ref{thm:main_result_matroids}. Since the pivot operation is
of order $2$, $\mathcal{M}_{G_1}*X = \mathcal{M}_{G_2}$. Hence,
$\mathcal{M}_{G_1}*X$ has a graph representation (graph $G_2$), and
thus $\emptyset$ is in set system $\mathcal{M}_{G_1}*X$.
Consequently, $X$ is in set system $\mathcal{M}_{G_1}$, thus $\det
G_1\sub{X} = 1$, and so $G_1*X = (G+X)*X$ is defined.
\end{Proof}
\fi

The reverse implication of Lemma~\ref{lem:def_left_right_pivots}
does not hold: take, e.g., $G$ to be the connected graph of two
vertices with each vertex having a loop.

We now state Theorem~\ref{thm:normal_form_pivot_loopc_matroids}
restricted to graphs.
\begin{Corollary} \label{cor:normal_form_pivot_loopc_graphs}
Let $G$ be a graph, and let $\varphi$ be a sequence of local, edge,
and loop complementation operations applicable to $G$. We have that
$G \varphi = G+X*Y+Z$ for some $X, Y, Z \subseteq V$ with $X
\subseteq Y$.
\end{Corollary}
\begin{Proof}
By Theorem~\ref{thm:normal_form_pivot_loopc_matroids},
$\mathcal{M}_{G} \varphi = \mathcal{M}_{G}+X*Y+Z$ for some $X, Y, Z
\subseteq V$ with $X \subseteq Y$. It suffices to show now that
$G+X*Y+Z$ is defined, i.e., show that $*Y$ is applicable to $G+X$.
As $\mathcal{M}_{G}+X*Y+Z$ represents a graph (the graph $G
\varphi$), $\mathcal{M}_{G}+X*Y$ also represents a graph (the graph
$G \varphi + Z$). Therefore, $\emptyset$ is in $\mathcal{M}_{G}+X*Y$
and thus $Y$ is in $\mathcal{M}_{G}+X$. Consequently, $*Y$ is indeed
applicable to $G+X$.
\end{Proof}

Corollary~\ref{cor:main_result_graphs} can also be proven directly
using Equality~(\ref{pivot_def_reverse}), i.e., the partial inverse
property of pivots. This is shown in the next theorem which also
provides a direct definition of the dual pivot for matrices.

Let $A$ be a $V \times V$-matrix and let $X \subseteq V$. We write
$A(x,y)^T$ to denote the application of $A$ to the vector $x \choose
y$, where it is understood that $x$ corresponds to the
$X$-coordinates, and $y$ to the remaining coordinates. We make now
an exception and consider arbitrary matrices, instead of symmetric
matrices, over $\two$. In this way the next result provides another
generalization (in addition to the generalization to set systems of
Theorem~\ref{thm:main_result_matroids}) of the concept of dual pivot
on graphs.

\begin{Theorem} \label{thm:main_result_graphs_alt}
Let $A$ be a $V \times V$-matrix over $\two$ and let $X \subseteq
V$. Then
$A+X*X+X = A*X+X*X$
(if both sides are defined), and
moreover $A(x_1,y_1)^T = (x_2,y_2)^T$ iff $(A+X*X+X)
(x_1+x_2,y_1)^T = (x_2,y_2)^T$ (if $A+X*X$ is defined). In
addition, any matrix $B$ with this property is of the form $B =
A+X*X+X$.
\end{Theorem}
\begin{Proof}
The pivot operation acts as a partial inverse, cf.
(\ref{pivot_def_reverse}). Hence $A(x_1,y_1)^T$ $= (x_2,y_2)^T$ iff
$(A*X)(x_2,y_1)^T = (x_1,y_2)^T$.
The loop complementation adds $1$ to the diagonal elements
corresponding to $X$, thus $A(x_1,y_1)^T = (x_2,y_2)^T$ iff
$(A+X)(x_1,y_1)^T = (x_1+x_2,y_2)^T$.

We simply chain these equalities: $A(x_1,y_1)^T = (x_2,y_2)^T$ iff
$(A+X)(x_1,y_1)^T$ $= (x_1+x_2,y_2)^T$ iff $(A+X*X)(x_1+x_2,y_1)^T =
(x_1,y_2)^T$ iff $(A+X*X+X) (x_1+x_2,y_1)^T = (x_2,y_2)^T$. We get a
similar result by chaining the equalities for $A*X+X*X$ instead of
$A+X*X+X$.

Finally, if a matrix $B$ exists with $B(x_1+x_2,y_1)^T =
(x_2,y_2)^T$ given the matrix $A$ with $A(x_1,y_1)^T = (x_2,y_2)^T$,
then $(B+X)(x_1 + x_2,y_1)^T = (x_1,y_2)^T$ and $(A+X)(x_1,y_1)^T =
(x_1 + x_2,y_2)^T$. Thus, by the definition of pivot given by
Equality~(\ref{pivot_def_reverse}) in Section~\ref{sec:def_pivots},
we have $A+X*X = B+X$, and so $B = A+X*X+X$.
\end{Proof}

It is interesting to consider
Theorem~\ref{thm:main_result_graphs_alt} for the case $X = V$.
Recall that for matrix $A$, $A*V$ is the inverse $A^{-1}$ of $A$.
Also, $A+V$ simply means adding the identity matrix (often denoted
by $I$) to $A$. Therefore, by
Theorem~\ref{thm:main_result_graphs_alt}, we see that over $\two$
addition of $I$ and matrix inversion together form the group $S_3$.
In particular, $((A^{-1}+I)^{-1}+I)^{-1}+I = A$ (assuming that the
left-hand side is defined).

\section{Maximal Pivots}

In this section we show that the dual pivot retains the maximal
elements $\max(M)$ (w.r.t.~inclusion) for any set system $M$, i.e.,
$\max(M)=\max(M \dual X)$ for any $X \subseteq V$. In this way we
generalize and provide an alternative proof for the main result of
\cite{MaxPivotsGraphs/Brijder09} where this result is shown for
graphs (i.e., the case $M = \mathcal{M}_G$):
$\max(\mathcal{M}_G)=\max(\mathcal{M}_{G \dual X})$ for graph $G$
and $X \subseteq V(G)$ such that $G \dual X$ is defined.

\begin{Remark}
More precisely, in \cite{MaxPivotsGraphs/Brijder09} the operation
$G+V*X+V$ is considered instead of $G \dual X = G+X*X+X$.
Now as pivot and loop complementation on disjoint sets commute (see
just below Example~\ref{ex:loop_compl_set_system}), $G+V*X+V =
G+X*X+X$ (as $V \setminus X$ and $X$ are disjoint, and the left-hand
side is defined iff the right-hand side is defined). Hence, this
operation is precisely the dual pivot $G\dual X$ restricted to
graphs $G$. In fact, $M \dual X$ defined in this paper is named
dual pivot as the corresponding graph operation $G+V*X+V$ in
\cite{MaxPivotsGraphs/Brijder09} is called dual pivot as well.
\end{Remark}

First we define the dual pivot explicitly for set systems. We have
$\alpha_{\bar *} =
\left( \begin{array}{cc} 1 & 0 \\
1 & 1 \end{array} \right)$. Hence, for $j \in V$, $M\dual \{j\} =
(V,D')$ where, for all $Z \subseteq V$, $Z \in D'$ iff
\begin{eqnarray} \label{def:dual_pivot_explicit}\begin{cases}
Z \in D & \mbox{if } j \in Z \\
(Z \cup \{j\} \in D) \xor (Z \in D) & \mbox{if } j \not\in Z \\
\end{cases}.\end{eqnarray}

Similarly as for loop complementation, we can reformulate the
definition of the dual pivot. If we let $M = (V,D)$, then
$M\dual \{j\} = (V,D')$ with $D' = D \xor \{Z \setminus \{j\} \mid j \in Z
\in D\}$. Moreover, we may provide a characterization of dual pivot
similar to the characterization of loop complementation in
Theorem~\ref{thm:loopc_ss_char}.

\begin{Theorem} \label{thm:dualp_ss_char}
Let $M$ be a set system and $X,Y \subseteq V$. We have $Y \in M\dual X$ iff $|\{ Z \in M \mid Y \subseteq Z \subseteq Y \cup X \}|$ is odd.
\end{Theorem}
\begin{Proof}
We apply Theorem~\ref{thm:loopc_ss_char} to $M \dual X =  M * V
+ X *V$ and use the fact that the operation ${} * V$
complements the sets of a set system. We have $Y \in M \dual X$
iff $V \setminus Y \in M * V + X$ iff the set $
\{ Z \in M * V \mid (V \setminus Y) \setminus X \subseteq Z \subseteq
V \setminus Y \}
=
\{ Z \in M  \mid (V \setminus Y) \setminus X \subseteq V \setminus Z
\subseteq V \setminus Y \}
=
\{ Z \in M  \mid Y \subseteq Z \subseteq X \cup Y \}
$
is of odd cardinality (where in the first equality we have
changed the variable $Z := Z \xor V$, and in the second
equality we applied $\xor V$ to invert both inclusions).
\end{Proof}

The following result is almost a direct consequence of
Theorems~\ref{thm:loopc_ss_char} and \ref{thm:dualp_ss_char}.
\begin{Theorem} \label{thm:max_pivots_dual}
Let $M$ be a set system over $V$ and $X \subseteq V$. Then
$\max(M)=\max(M \dual X)$ and $\min(M)=\min(M + X)$.
\end{Theorem}
\begin{Proof}
If $Y \in \max(M)$, then $Y \in M\dual X$ by
Theorem~\ref{thm:dualp_ss_char} (as $\{ Z \in M \mid Y
\subseteq Z \subseteq Y \cup X \} = \{Y\}$). Let $M' = M \dual
X$. By exactly the same reasoning as before, we find that $Y \in
\max(M')$ implies that $Y \in M'\dual X = M$. Hence
$\max(M)=\max(M \dual X)$.

Similarly, the equality $\min(M)=\min(M + X)$ follows from
Theorem~\ref{thm:loopc_ss_char}.
\end{Proof}
%
%

\begin{Example} \label{ex:max_pivots_set_system}
Let $V = \{p,q,r,s\}$ and $M = (V,D)$ with
$$D = \{\varnothing, \{p\}, \{q\}, \{p,r\}, \{p,s\}, \{r,s\},
\{p,q,r\}, \{p,q,s\}, \{p,r,s\}, \{q,r,s\}\}.$$
Then $M\dual \{r\} = (V,D')$ with
$$
D' = \{\varnothing,  \{q\}, \{s\}, \{p,q\}, \{p,r\}, \{q,s\},
\{r,s\}, \{p,q,r\}, \{p,q,s\}, \{p,r,s\}, \{q,r,s\}\}.
$$
Thus indeed $\max(M) = \{\{p,q,r\}, \{p,q,s\}, \{p,r,s\},
\{q,r,s\}\} =$ $\max(M\dual \{r\})$. Note that the maximal elements
may differ dramatically when performing (regular) pivot or loop
complementation: e.g., $\max(M*\{q\}) = \{\{p,q,r,s\}\}$.
\end{Example}

The corresponding result restricted to graphs is given below for
completeness. The result is shown in
\cite{MaxPivotsGraphs/Brijder09} using linear algebra techniques,
while in this paper it is almost a direct consequence of the
definition of dual pivot on set systems. Note that for graph $G$ and
$X \subseteq V(G)$, $X \in \max(\mathcal{M}_G)$ iff both $\det
G\sub{X} = 1$ and $\det G\sub{Y} = 0$ for every $Y \supset X$.

\begin{Corollary}[\cite{MaxPivotsGraphs/Brijder09}] \label{cor:max_pivots_dual}
Let $G$ be a graph, and let $X \subseteq V(G)$. Then
$\max(\mathcal{M}_G) = \max(\mathcal{M}_{G \dual X})$ if the
right-hand side is defined (i.e., $\det (G+X)\sub{X} = 1$).
\end{Corollary}

While the result $\min(M)=\min(M + X)$
(in Theorem~\ref{thm:max_pivots_dual}) may be relevant
for arbitrary set systems, the result is trivial when restricted to
graphs. Indeed, for a graph $G$ we have $\min(\mathcal{M}_G) =
\{\emptyset\}$ and since $\mathcal{M}_G+X$ represents a graph (it is
the graph $G+X$) we have $\min(\mathcal{M}_G+X) = \{\emptyset\}$.

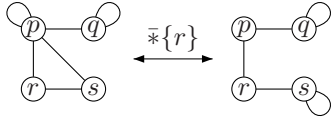
\begin{figure}[t]
\unitlength 0.7mm%
\begin{center}
%
\begin{picture}(55,17)(-10,-05)
\node[Nw=25,Nh=25,Nframe=n](I)(00,00){%
  \unitlength0.4mm
  \begin{picture}(30,30)
  \pivotorbitmacro
  \drawedge(p,q){}
  \drawedge(p,r){}
  \drawedge(p,s){}
  \drawedge(r,s){}
  \drawloop[loopangle=45,loopdiam=7](q){}
  \drawloop[loopangle=135,loopdiam=7](p){}
  \end{picture}
}
\node[Nw=25,Nh=25,Nframe=n](II)(40,00){%
  \unitlength0.4mm
  \begin{picture}(30,30)
  \pivotorbitmacro
  \drawedge(p,q){}
  \drawedge(p,r){}
  \drawedge(r,s){}
  \drawloop[loopangle=45,loopdiam=7](q){}
  \drawloop[loopangle=-45,loopdiam=7](s){}
\end{picture}
}
  \drawedge[AHnb=1,ATnb=1](I,II){$\dual \{r\}$}
\end{picture}
\end{center}
\caption{Dual pivot $\dual \{r\}$ on graph $G$ from the upper-left
corner of Figure~\ref{fig:pivot_space}.} \label{fig:dual_pivot_ex}
\end{figure}

\begin{Example} \label{ex:max_pivots_graph}
Set system $M$ of Example~\ref{ex:max_pivots_set_system} corresponds
to graph $G$ on the upper-left corner of
Figure~\ref{fig:pivot_space}. For $X = \{r\}$, $\det (G+X)\sub{X} =
1$ holds as $\{r\}$ is a loop in $G+\{r\}$. Graphs $G$ and
$G\dual \{r\}$ are given in Figure~\ref{fig:dual_pivot_ex}.
\end{Example}

The proof of Corollary~\ref{cor:max_pivots_dual} in
\cite{MaxPivotsGraphs/Brijder09} relies heavily on the fact that the
elements of $\max(\mathcal{M}_G)$ are all of cardinality equal to
the rank of (the adjacency matrix of) $G$, a consequence of the
Strong Principal Minor Theorem, see
\cite[Theorem~2.9]{Kodiyalam_Lam_Swan_2008}. This property of
$\max(\mathcal{M}_G)$ turns out to be irrelevant for
Corollary~\ref{cor:max_pivots_dual} as its generalization,
Theorem~\ref{thm:max_pivots_dual}, holds for set systems in general
where this property of $\max(\mathcal{M}_G)$ of course does not
hold.

In \cite{MaxPivotsGraphs/Brijder09} it was also noted that the
kernel (null space) of a graph is invariant under dual pivot.
It is straightforward to verify now using
Theorem~\ref{thm:main_result_graphs_alt} that this holds for
arbitrary matrices over $\two$: if $A(x_1,y_1)^T = (0,0)$, then
$A\dual X (x_1+0,y_1)^T = (0,0)$. Therefore, $A\dual X
(x_1,y_1)^T = (0,0)$. The converse holds as dual pivot is an
involution (operation of order $2$). In particular, the rank of
$A$ is invariant w.r.t.~the dual pivot.

As observed in \cite{MaxPivotsGraphs/Brijder09}, as a graph
transformation operation, the dual pivot is similar to the
(regular) pivot. More precisely, the elementary dual pivots
$G\dual X$ are either of the form (1) $X = \{u\}$ where $u$
does not have a loop in $G$ or of the form (2) $X = \{u,v\}$
where $\{u,v\}$ is an edge of $G$ where both $u$ and $v$ have
loops. The effect of elementary pivot $\dual \{u\}$ is the same
as that of $*\{u\}$, complementing its neighbourhood. Similarly for
elementary pivot $\dual \{u,v\}$. Only the conditions for
applying of elementary dual pivots are different compared to those for
(regular) elementary pivots: the effect of the operation is the
same.

\section{Consequences for Simple Graphs}

In this section we consider simple graphs, i.e., undirected
graphs without loops or parallel edges. Local complementation
was first studied on simple graphs \cite{kotzig1968}: local
complementation on a vertex $u$, by abuse of notation denoted
by $*\{u\}$, complements the edges in the neighbourhood of $u$,
thus it is the same operation as for graphs (loops allowed)
except that applicability is not dependent on the presence of a
loop on $u$, and neither are loops added or removed in the
neighbourhood. Also edge complementation $*\{u,v\}$ on edge
$\{u,v\}$ for simple graphs is defined as for graphs, inverting
certain sets of edges, cf. Figure~\ref{fig:pivot}, but again
the absence of loops is not an (explicit) requirement for
applicability.

The ``curious'' identity ${}*\{u,v\} = {}*\{u\}*\{v\}*\{u\}$ for simple
graphs shown by Bouchet~\cite[Corollary~8.2]{bouchet1988} and found
in standard textbooks, see, e.g.,
\cite[Theorem~8.10.2]{AGT_godsil_2001}, can be proven by a
straightforward (but slightly tedious) case analysis involving $u$,
$v$ and all possible combinations of their neighbours. Here it is
obtained, cf. Proposition~\ref{prop:class_uvu_vuv}, as a consequence
of Theorem~\ref{thm:main_result_matroids}.

\newcommand{\yhoog}{22}
\newcommand{\yhooghalf}{11}
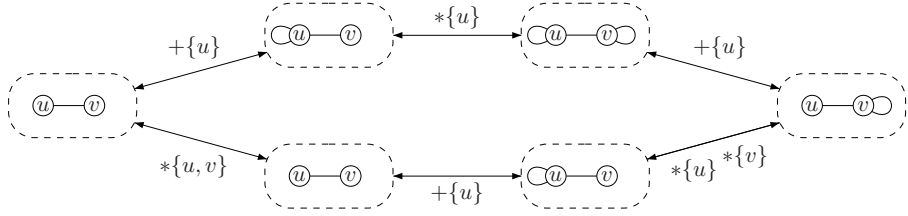
\begin{figure}[t]
\unitlength 1mm%
\begin{center}
\resizebox{\textwidth}{!}{ 
%
\begin{picture}(140,30)(-10,-05)
%
\node[Nw=20,Nh=10,Nframe=y,dash={1}0](g)(0,\yhooghalf){}
\node[Nw=20,Nh=10,Nframe=n](qs)(0,\yhooghalf){%
  \unitlength0.4mm
  \begin{picture}(30,10)
  \gasset{AHnb=0,Nw=1.5,Nh=1.5,Nframe=n,Nfill=y}
  \gasset{AHnb=0,Nw=8,Nh=8,Nframe=y,Nfill=n}
  \node(1)(05,05){$u$}
  \node(2)(25,05){$v$}
  \drawedge(1,2){}
\end{picture}
}
\node[Nw=20,Nh=10,Nframe=y,dash={1}0](gp)(40,\yhoog){}
\node[Nw=20,Nh=10,Nframe=n](rs)(40,\yhoog)%
{%
  \unitlength0.4mm
  \begin{picture}(30,10)
  \gasset{AHnb=0,Nw=1.5,Nh=1.5,Nframe=n,Nfill=y}
  \gasset{AHnb=0,Nw=8,Nh=8,Nframe=y,Nfill=n}
  \node(1)(05,05){$u$}
  \node(2)(25,05){$v$}
  \drawedge(1,2){}
  \drawloop[loopangle=180,loopdiam=7](1){}
\end{picture}
}
\node[Nw=20,Nh=10,Nframe=y,dash={1}0](gs)(40,0){}
\node[Nw=20,Nh=10,Nframe=n](pr)(40,0){%
  \unitlength0.4mm
  \begin{picture}(30,10)
  \gasset{AHnb=0,Nw=1.5,Nh=1.5,Nframe=n,Nfill=y}
  \gasset{AHnb=0,Nw=8,Nh=8,Nframe=y,Nfill=n}
  \node(1)(05,05){$u$}
  \node(2)(25,05){$v$}
  \drawedge(1,2){}
\end{picture}
}
\node[Nw=20,Nh=10,Nframe=y,dash={1}0](gps)(80,\yhoog){}
\node[Nw=20,Nh=10,Nframe=n](qr)(80,\yhoog){%
  \unitlength0.4mm
  \begin{picture}(30,10)
  \gasset{AHnb=0,Nw=1.5,Nh=1.5,Nframe=n,Nfill=y}
  \gasset{AHnb=0,Nw=8,Nh=8,Nframe=y,Nfill=n}
  \node(1)(05,05){$u$}
  \node(2)(25,05){$v$}
  \drawedge(1,2){}
  \drawloop[loopangle=180,loopdiam=7](1){}
  \drawloop[loopangle=0,loopdiam=7](2){}
\end{picture}
}
\node[Nw=20,Nh=10,Nframe=y,dash={1}0](gsp)(80,0){}
\node[Nw=20,Nh=10,Nframe=n](pq)(80,0){%
  \unitlength0.4mm
  \begin{picture}(30,10)
  \gasset{AHnb=0,Nw=1.5,Nh=1.5,Nframe=n,Nfill=y}
  \gasset{AHnb=0,Nw=8,Nh=8,Nframe=y,Nfill=n}
  \node(1)(05,05){$u$}
  \node(2)(25,05){$v$}
  \drawedge(1,2){}
  \drawloop[loopangle=180,loopdiam=7](1){}
\end{picture}
}
\node[Nw=20,Nh=10,Nframe=y,dash={1}0](gpsp)(120,\yhooghalf){}
\node[Nw=20,Nh=10,Nframe=n](ps)(120,\yhooghalf){%
  \unitlength0.4mm
  \begin{picture}(30,10)
  \gasset{AHnb=0,Nw=1.5,Nh=1.5,Nframe=n,Nfill=y}
  \gasset{AHnb=0,Nw=8,Nh=8,Nframe=y,Nfill=n}
  \node(1)(05,05){$u$}
  \node(2)(25,05){$v$}
  \drawedge(1,2){}
  \drawloop[loopangle=0,loopdiam=7](2){}
\end{picture}
}
\drawedge[AHnb=1,ATnb=1](g,gp){$+\{u\}$}
\drawedge[AHnb=1,ATnb=1,ELside=r](g,gs){$*\{u,v\}$}
\drawedge[AHnb=1,ATnb=1](gp,gps){$*\{u\}$}
\drawedge[AHnb=1,ATnb=1,ELside=r](gs,gsp){$+\{u\}$}
\drawedge[AHnb=1,ATnb=1](gps,gpsp){$+\{u\}$}
\drawedge[AHnb=1,ATnb=1,ELpos=40,ELside=r](gsp,gpsp){$*\{u\}$}
\drawedge[AHnb=0,ATnb=0,ELpos=60,ELside=r](gsp,gpsp){$*\{v\}$}
\end{picture}
}
\end{center}
\caption{Verification of applicability of $*\{u,v\} +\{u\} *\{u\}
*\{v\} +\{u\} *\{u\} +\{u\}$ to any graph $F$ having an edge $\{u,v\}$
with both $u$ and $v$ non-loop vertices.}
\label{fig:local_compl_uvu_uv}
\end{figure}

\begin{Proposition}
\label{prop:class_uvu_vuv} Let $H$ be a simple graph having an
edge $\{u,v\}$. We have $H *\{u,v\} = H *\{u\}*\{v\}*\{u\} = H
*\{v\}*\{u\}*\{v\}$.
\end{Proposition}
\begin{Proof}
Let $M$ be a set system, and $u$ and $v$ two distinct elements
from its domain. Define $\varphi = *\{u,v\} +\{u\} *\{u\}
*\{v\} +\{u\} *\{u\} +\{u\}$. Recall that for set systems we
have $*\{u,v\} = *\{u\} *\{v\}$ and that the operations on
different elements commute, e.g. $*\{v\}+\{u\} = +\{u\}*\{v\}$.
We have therefore $\varphi = *\{u\} *\{v\} +\{u\} *\{u\} *\{v\}
+\{u\} *\{u\} +\{u\} =  *\{u\} +\{u\} *\{u\} +\{u\} *\{u\}
+\{u\} = \mathrm{id}$, where in the last equality we used
Theorem~\ref{thm:main_result_matroids}. Therefore, $M \varphi =
M$ for any set system $M$ having $u$ and $v$ in its domain.

Hence, any graph $G$ for which $\varphi$ is applicable to $G$,
we have $G \varphi = G$. Assume now that $G$ is a graph
(allowing loops) having the edge $\{u,v\}$ where both $u$ and
$v$ do not have a loop. By Figure~\ref{fig:local_compl_uvu_uv}
we see that $\varphi$ is applicable to $G$, and therefore $G
\varphi = G$.

Now, modulo loops, i.e., considering simple graphs $H$, we no
longer worry about the presence of loops, and we may omit the
loop complementation operations from $\varphi$. Hence $*\{u,v\}
*\{u\} *\{v\} *\{u\}$ is the identity on simple graphs, and
therefore $*\{u,v\} = *\{u\} *\{v\} *\{u\}$. By symmetry of the
$*\{u,v\}$ operation we also have that $*\{u,v\} = *\{v\}
*\{u\} *\{v\}$.
\end{Proof}

Thus, for set systems we have the decomposition $*\{u,v\} =
*\{u\} *\{v\}$, whereas for simple graphs the decomposition of
edge complementation into local complementation takes the form
$*\{u,v\} = *\{u\} *\{v\} *\{u\}$. The rationale behind this
last equality is hidden, as in fact the equality $*\{u,v\} =
+\{u\} *\{u\} +\{u\} *\{v\} *\{u\} +\{u\}$ is demonstrated for
graphs (loops allowed) (see the proof of
Proposition~\ref{prop:class_uvu_vuv}). The fact that the
equality of Proposition~\ref{prop:class_uvu_vuv} does not hold
for graphs (with loops allowed) is a consequence of the added
requirement of applicability of the operations. Applicability
depends on the presence or absence of loops, and it is curious
that loops are necessary to fully understand these operations
for simple graphs (which are loopless by definition)!

A second remark concerns
Figure~\ref{fig:local_compl_uvu_uv} and its role in the proof.
Following the operations around the loop in the diagram, starting
and ending at the same point, we obtain the identity operation (on
set systems). The diagram in the figure does not show that the
identity holds, it merely concerns \emph{applicability} of the
operations (in graphs). It is possible to graphically verify that
composing the operations around the loop forms the identity: one
has to add several ``generic'' vertices $q$ each representing a
specific case of whether or not $u$ and whether or not $v$ is in the
neighbourhood of $q$. However, the number of vertices $q$ grows
exponentially in the number of vertices of the subgraph (in this
case an edge consisting of vertices $u$ and $v$) under
consideration. Here, verifying the applicability of $\varphi$ on the
subgraph induced by $u$ and $v$ suffices.

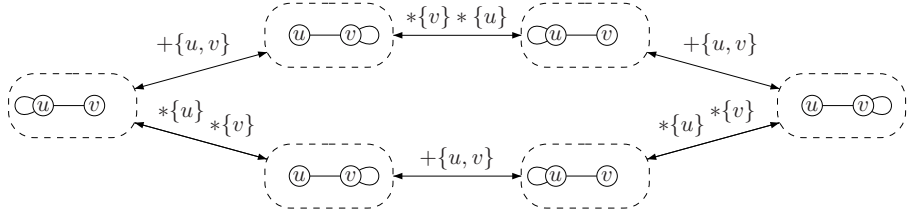
\begin{figure}[t]
\unitlength 1mm%
\begin{center}
\resizebox{\textwidth}{!}{ 
%
\begin{picture}(140,30)(-10,-05)
\node[Nw=20,Nh=10,Nframe=y,dash={1}0](g)(0,\yhooghalf){}
\node[Nw=20,Nh=10,Nframe=n](qs)(0,\yhooghalf){%
  \unitlength0.4mm
  \begin{picture}(30,10)
  \gasset{AHnb=0,Nw=1.5,Nh=1.5,Nframe=n,Nfill=y}
  \gasset{AHnb=0,Nw=8,Nh=8,Nframe=y,Nfill=n}
  \node(1)(05,05){$u$}
  \node(2)(25,05){$v$}
  \drawedge(1,2){}
  \drawloop[loopangle=180,loopdiam=7](1){}
\end{picture}
}
\node[Nw=20,Nh=10,Nframe=y,dash={1}0](gp)(40,\yhoog){}
\node[Nw=20,Nh=10,Nframe=n](rs)(40,\yhoog){%
  \unitlength0.4mm
  \begin{picture}(30,10)
  \gasset{AHnb=0,Nw=1.5,Nh=1.5,Nframe=n,Nfill=y}
  \gasset{AHnb=0,Nw=8,Nh=8,Nframe=y,Nfill=n}
  \node(1)(05,05){$u$}
  \node(2)(25,05){$v$}
  \drawedge(1,2){}
  \drawloop[loopangle=0,loopdiam=7](2){}
\end{picture}
}
\node[Nw=20,Nh=10,Nframe=y,dash={1}0](gs)(40,0){}
\node[Nw=20,Nh=10,Nframe=n](pr)(40,0){%
  \unitlength0.4mm
  \begin{picture}(30,10)
  \gasset{AHnb=0,Nw=1.5,Nh=1.5,Nframe=n,Nfill=y}
  \gasset{AHnb=0,Nw=8,Nh=8,Nframe=y,Nfill=n}
  \node(1)(05,05){$u$}
  \node(2)(25,05){$v$}
  \drawedge(1,2){}
  \drawloop[loopangle=0,loopdiam=7](2){}
\end{picture}
}
\node[Nw=20,Nh=10,Nframe=y,dash={1}0](gps)(80,\yhoog){}
\node[Nw=20,Nh=10,Nframe=n](qr)(80,\yhoog){%
  \unitlength0.4mm
  \begin{picture}(30,10)
  \gasset{AHnb=0,Nw=1.5,Nh=1.5,Nframe=n,Nfill=y}
  \gasset{AHnb=0,Nw=8,Nh=8,Nframe=y,Nfill=n}
  \node(1)(05,05){$u$}
  \node(2)(25,05){$v$}
  \drawedge(1,2){}
  \drawloop[loopangle=180,loopdiam=7](1){}
\end{picture}
}
\node[Nw=20,Nh=10,Nframe=y,dash={1}0](gsp)(80,0){}
\node[Nw=20,Nh=10,Nframe=n](pq)(80,0){%
  \unitlength0.4mm
  \begin{picture}(30,10)
  \gasset{AHnb=0,Nw=1.5,Nh=1.5,Nframe=n,Nfill=y}
  \gasset{AHnb=0,Nw=8,Nh=8,Nframe=y,Nfill=n}
  \node(1)(05,05){$u$}
  \node(2)(25,05){$v$}
  \drawedge(1,2){}
  \drawloop[loopangle=180,loopdiam=7](1){}
\end{picture}
}
\node[Nw=20,Nh=10,Nframe=y,dash={1}0](gpsp)(120,\yhooghalf){}
\node[Nw=20,Nh=10,Nframe=n](ps)(120,\yhooghalf){%
  \unitlength0.4mm
  \begin{picture}(30,10)
  \gasset{AHnb=0,Nw=1.5,Nh=1.5,Nframe=n,Nfill=y}
  \gasset{AHnb=0,Nw=8,Nh=8,Nframe=y,Nfill=n}
  \node(1)(05,05){$u$}
  \node(2)(25,05){$v$}
  \drawedge(1,2){}
  \drawloop[loopangle=0,loopdiam=7](2){}
\end{picture}
}
\drawedge[AHnb=1,ATnb=1](g,gp){$+\{u,v\}$}
\drawedge[AHnb=1,ATnb=1,ELpos=40](g,gs){$*\{u\}$}
\drawedge[AHnb=0,ATnb=0,ELpos=60](g,gs){$*\{v\}$}
\drawedge[AHnb=1,ATnb=1](gp,gps){$*\{v\}*\{u\}$}
\drawedge[AHnb=1,ATnb=1](gs,gsp){$+\{u,v\}$}
\drawedge[AHnb=1,ATnb=1](gps,gpsp){$+\{u,v\}$}
\drawedge[AHnb=1,ATnb=1,ELpos=40](gsp,gpsp){$*\{u\}$}
\drawedge[AHnb=0,ATnb=0,ELpos=60](gsp,gpsp){$*\{v\}$}
\end{picture}
} 
\end{center}
\caption{Verification of applicability of $(*\{u\}*\{v\}
+\{u,v\})^3$ to a graph $G$ having an edge $\{u,v\}$ with a loop on
vertex $u$.} \label{fig:local_compl_uvu}
\end{figure}

Incidentally, the equality $*\{u\}*\{v\}*\{u\} =
*\{v\}*\{u\}*\{v\}$ can also be verified directly by using
Figure~\ref{fig:local_compl_uvu} instead of
Figure~\ref{fig:local_compl_uvu_uv} in the proof of
Proposition~\ref{prop:class_uvu_vuv}, and observing that that
$(*\{u\}*\{v\} +\{u,v\})^3$  is the identity (in set systems). This
does not show the equality to $*\{u,v\}$ in simple graphs.

\begin{figure}[t]
\unitlength 1mm%
\begin{center}
\resizebox{\textwidth}{!}{ 
\begin{picture}(140,55)(-10,-05)
\node[Nw=20,Nh=20,Nframe=y,dash={1}0](g)(0,20){}
\node[Nw=20,Nh=20,Nframe=n](qs)(0,20){%
  \unitlength0.4mm
  \begin{picture}(30,10)
  \gasset{AHnb=0,Nw=1.5,Nh=1.5,Nframe=n,Nfill=y}
  \gasset{AHnb=0,Nw=8,Nh=8,Nframe=y,Nfill=n}
  \node(1)(05,05){$u$}
  \node(2)(25,18){$v$}
  \node(3)(25,-8){$w$}
  \drawedge(1,2){}
  \drawedge(1,3){}
  \drawedge(2,3){}
  \drawloop[loopangle=180,loopdiam=7](1){}
\end{picture}
}
\node[Nw=20,Nh=20,Nframe=y,dash={1}0](gp)(40,40){}
\node[Nw=20,Nh=20,Nframe=n](rs)(40,40){%
  \unitlength0.4mm
  \begin{picture}(30,10)
  \gasset{AHnb=0,Nw=1.5,Nh=1.5,Nframe=n,Nfill=y}
  \gasset{AHnb=0,Nw=8,Nh=8,Nframe=y,Nfill=n}
  \node(1)(05,05){$u$}
  \node(2)(25,18){$v$}
  \node(3)(25,-8){$w$}
  \drawedge(1,2){}
  \drawedge(1,3){}
  \drawedge(2,3){}
  \drawloop[loopangle=180,loopdiam=7](1){}
  \drawloop[loopangle=0,loopdiam=7](2){}
\end{picture}
}
\node[Nw=20,Nh=20,Nframe=y,dash={1}0](gs)(40,0){}
\node[Nw=20,Nh=20,Nframe=n](pr)(40,0){%
  \unitlength0.4mm
  \begin{picture}(30,10)
  \gasset{AHnb=0,Nw=1.5,Nh=1.5,Nframe=n,Nfill=y}
  \gasset{AHnb=0,Nw=8,Nh=8,Nframe=y,Nfill=n}
  \node(1)(05,05){$u$}
  \node(2)(25,18){$v$}
  \node(3)(25,-8){$w$}
  \drawedge(1,2){}
  \drawedge(1,3){}
  \drawloop[loopangle=180,loopdiam=7](1){}
  \drawloop[loopangle=0,loopdiam=7](2){}
  \drawloop[loopangle=0,loopdiam=7](3){}
\end{picture}
}
\node[Nw=20,Nh=20,Nframe=y,dash={1}0](gps)(80,40){}
\node[Nw=20,Nh=20,Nframe=n](qr)(80,40){%
  \unitlength0.4mm
  \begin{picture}(30,10)
  \gasset{AHnb=0,Nw=1.5,Nh=1.5,Nframe=n,Nfill=y}
  \gasset{AHnb=0,Nw=8,Nh=8,Nframe=y,Nfill=n}
  \node(1)(05,05){$u$}
  \node(2)(25,18){$v$}
  \node(3)(25,-8){$w$}
  \drawedge(1,2){}
  \drawedge(2,3){}
  \drawloop[loopangle=0,loopdiam=7](2){}
  \drawloop[loopangle=0,loopdiam=7](3){}
\end{picture}
}
\node[Nw=20,Nh=20,Nframe=y,dash={1}0](gsp)(80,0){}
\node[Nw=20,Nh=20,Nframe=n](pq)(80,0){%
  \unitlength0.4mm
  \begin{picture}(30,10)
  \gasset{AHnb=0,Nw=1.5,Nh=1.5,Nframe=n,Nfill=y}
  \gasset{AHnb=0,Nw=8,Nh=8,Nframe=y,Nfill=n}
  \node(1)(05,05){$u$}
  \node(2)(25,18){$v$}
  \node(3)(25,-8){$w$}
  \drawedge(1,2){}
  \drawedge(1,3){}
  \drawloop[loopangle=180,loopdiam=7](1){}
  \drawloop[loopangle=0,loopdiam=7](3){}
\end{picture}
}
\node[Nw=20,Nh=20,Nframe=y,dash={1}0](gpsp)(120,20){}
\node[Nw=20,Nh=20,Nframe=n](ps)(120,20){%
  \unitlength0.4mm
  \begin{picture}(30,10)
  \gasset{AHnb=0,Nw=1.5,Nh=1.5,Nframe=n,Nfill=y}
  \gasset{AHnb=0,Nw=8,Nh=8,Nframe=y,Nfill=n}
  \node(1)(05,05){$u$}
  \node(2)(25,18){$v$}
  \node(3)(25,-8){$w$}
  \drawedge(1,2){}
  \drawedge(2,3){}
  \drawloop[loopangle=0,loopdiam=7](3){}
\end{picture}
}
\drawedge[AHnb=1,ATnb=1](g,gp){$+\{v\}$}
\drawedge[AHnb=1,ATnb=1,ELpos=35](g,gs){$*\{u\}$}
\drawedge[AHnb=0,ATnb=0,ELpos=50](g,gs){$*\{v\}$}
\drawedge[AHnb=0,ATnb=0,ELpos=65](g,gs){$*\{w\}$}
\drawedge[AHnb=1,ATnb=1](gp,gps){$*\{v\}$}
\drawedge[AHnb=1,ATnb=1](gs,gsp){$+\{v\}$}
\drawedge[AHnb=1,ATnb=1](gps,gpsp){$+\{v\}$}
\drawedge[AHnb=1,ATnb=1,ELpos=35](gsp,gpsp){$*\{u\}$}
\drawedge[AHnb=0,ATnb=0,ELpos=50](gsp,gpsp){$*\{v\}$}
\drawedge[AHnb=0,ATnb=0,ELpos=65](gsp,gpsp){$*\{w\}$}
\end{picture}
} 
\end{center}
\caption{Verification of applicability of a sequence of local and
loop complementations from
Corollary~\ref{cor:triangle_id_simple_graphs} to a graph $G$ where
$G\sub{\{u,v,w\}}$ is the left-most graph in the figure.}
\label{fig:new_eqn_simple_graphs}
\end{figure}
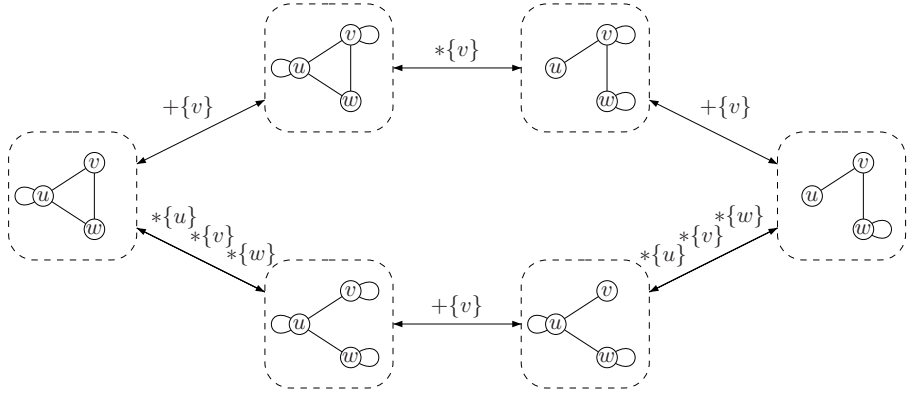

In addition to providing a new proof for
Proposition~\ref{prop:class_uvu_vuv}, the presented method
allows one to obtain many more curious equalities involving
local complementation and/or edge complementation. The steps
are as follows. One starts with an identity for set systems,
involving pivot and loop complementation. Then one shows
applicability for (general) graphs for the sequence of
operations. Finally one drops the loop complementation
operations to obtain an identity for simple graphs.

We illustrate this by stating one such equality.
Proposition~\ref{prop:class_uvu_vuv} considers the case where
$u,v \in V(H)$ is such that the subgraph of $H$ induced by
$\{u,v\}$ is a complete graph (i.e., $\{u,v\}$ is an edge in
$H$). We now deduce an equality where three vertices induce a
complete graph.

\begin{Corollary}
\label{cor:triangle_id_simple_graphs} Let $H$ be a simple
graph, and let $u,v,w \in V(H)$ be such that the subgraph of
$H$ induced by $\{u,v,w\}$ is a complete graph. Then $H
(*\{u\}*\{v\}*\{w\})^2 = H *\{v\}$.
\end{Corollary}
\ifshort\else
\begin{Proof}
The proof of this lemma is very similar to the proof of
Proposition~\ref{prop:class_uvu_vuv}. We have
$*\{u\}*\{v\}*\{w\}+\{v\}*\{u\}*\{v\}*\{w\} = *\{v\}+\{v\}*\{v\}$ as
pivot and loop complementation on disjoint sets commute. Moreover,
$*\{v\}+\{v\}*\{v\} = +\{v\}*\{v\}+\{v\}$ by
Theorem~\ref{thm:main_result_matroids}.

By Figure~\ref{fig:new_eqn_simple_graphs} we see that both
$*\{u\}*\{v\}*\{w\}+\{v\}*\{u\}*\{v\}*\{w\}$ and
$+\{v\}*\{v\}+\{v\}$ are applicable to any graph $G$ where
$G\sub{\{u,v,w\}}$ (the left-most graph in the figure) has loop
$\{u\}$ and edges $\{u,v\}$, $\{u,w\}$, and $\{v,w\}$. The
result follows by considering the equality modulo loops, i.e.,
``forgetting'' about loops.
\end{Proof} \fi

\begin{Remark}
Sabidussi \cite{DBLP:journals/dm/Sabidussi87} studies local
complementation on simple graphs with bicoloured vertices.
Local complementation on a vertex $u$ then also toggles the
colours of the vertices adjacent to $u$. By modelling the two
colours by the existence or nonexistence of loops, we find that
this operation is exactly local complementation in graphs,
where we additionally allow local complementation to be applied
on non-looped vertices. Let us denote this operation on a
vertex $u$ by $\tilde{*}\{u\}$. Hence, $\tilde{*}\{u\}$ is
equal to $*\{u\}$ if $u$ has a loop and equal to $\dual \{u\}$
if $u$ has a no loop.

In this context, we may reconsider the equality $G(*\{u\}*\{v\}
+\{u,v\})^3 = G$ from Figure~\ref{fig:local_compl_uvu} where
$G$ has an edge $\{u,v\}$ with $u$ and $v$ non-looped vertices.
We have that $\tilde{*}\{u\}$ and $+\{u\}$ commute as a loop is
of no consequence for applicability of $\tilde{*}$ (or more
formally, as $*\{u\}+\{u\} = +\{u\}\dual\{u\}$). We infer that
$G(\tilde{*}\{u\}\tilde{*}\{v\})^3= G+\{u,v\}$, and obtain in
this way \cite[Lemma~1]{DBLP:journals/dm/Sabidussi87}.

Similarly the equality
$G*\{u\}*\{v\}*\{w\}+\{v\}*\{u\}*\{v\}*\{w\} =
G+\{v\}*\{v\}+\{v\}$ where $G$ has a triangle, as proved in
Corollary~\ref{cor:triangle_id_simple_graphs}, see
Figure~\ref{fig:new_eqn_simple_graphs}, reduces to
$G\tilde{*}\{v\}(\tilde{*}\{w\}\tilde{*}\{v\}\tilde{*}\{u\})^2
= G+\{v\}$. Thus we also have obtained in this way
\cite[Lemma~2]{DBLP:journals/dm/Sabidussi87}.

Together these two results form the core of the central result
in \cite{DBLP:journals/dm/Sabidussi87} that any bicoloured
simple graph may be colour reversed by a linear number of local
complementation operations. Equivalently, $G+V$ can be obtained
from $G$ by a sequence of $\tilde{*}$ operations (of length
linear in $|V|$).
\end{Remark}

In the next result, Theorem~\ref{thm:normal_form_seq_loc}, we
go back-and-forth between the notions of simple graph and
graph. To avoid confusion, we explicitly formalize these
transitions. For a simple graph $H$, we define $i(H)$ to be $H$
regarded as a \emph{graph} (i.e., symmetric matrix over $\two$)
having no loops. Similarly, for graph $G$, we define $\pi(G)$
to be the simple graph obtained from $G$ by removing the loops.
Thus, $i(H)$ is the obvious injection from the set of simple
graphs to the set of graphs, while $\pi(G)$ is the obvious
projection from the set of graphs to the set of simple graphs.
We will use the following identities.
\begin{Lemma} \label{lem:proj_inj}
For simple graph $H$, $\pi(i(H)) = H$. For graph $G$ and
elementary pivot $G*X$ (hence $*X$ is either local or edge
complementation), $\pi(G*X) = \pi(G)*X$. Moreover, for $Y
\subseteq V(G)$, $\pi(G+Y) = \pi(G)$.
\end{Lemma}

If $\varphi$ is a sequence of edge complementation operations
applicable to graph $G$, then $\varphi(G) = G*Y$ for some $Y
\subseteq V(G)$, see \cite{BHH/PivotsDetPM/09} (or
alternatively, it may deduced from
\cite[Section~2]{DBLP:journals/ejc/Bouchet01},
\cite{bouchet1987}, and observing that the matrix operation
considered in these papers is, modulo $\two$, equal to
principal pivot transform). The converse also holds: if graph
$G$ does not have loops, then $G*Y$ is applicable iff $Y$ can
be decomposed into a sequence of applicable edge
complementation operations (i.e., all elementary pivot
operations are edge complementations). Similarly, as a
consequence of
Theorem~\ref{thm:normal_form_pivot_loopc_matroids}, the
following result characterizes the effect of sequences of local
complementations on simple graphs.

\begin{Theorem}
\label{thm:normal_form_seq_loc} Let $H$ be a simple graph, and
let $\varphi$ be a sequence of local complementation operations
applicable to $H$. Then $H\varphi = \pi(i(H)+X*Y)$ for some $X,
Y \subseteq V$ with $X \subseteq Y$.

Conversely, for graph $G$, if $G+X*Y$ is defined for some $X, Y
\subseteq V$, then there is a sequence $\varphi$ of local
complementation operations applicable to $\pi(G)$ such that
$\pi(G)\varphi = \pi(G+X*Y)$.
\end{Theorem}
\ifshort\else
\begin{Proof}
We first prove the first statement of the theorem. Let $\varphi
= *\{v_1\} \cdots *\{v_n\}$. We have, for any graph $G$ and $u
\in V(G)$, either $G*\{u\}$ is defined or $G+\{u\}*\{u\}$ is
defined (but not both). Thus there is a (unique) $\varphi' =
\varphi'_1 \varphi'_2 \cdots \varphi'_n$, where $\varphi'_i$ is
either $*\{v_i\}$ or $+\{v_i\}*\{v_i\}$ for all $i \in
\{1,\ldots,n\}$, such that $\varphi'$ is defined on (applicable
to) $i(H)$. By
Corollary~\ref{cor:normal_form_pivot_loopc_graphs},
$i(H)\varphi' = i(H)+X*Y+Z$ for some $X, Y, Z \subseteq V$ with
$X \subseteq Y$. By Lemma~\ref{lem:proj_inj}, $\pi(i(H)+X*Y) =
\pi(i(H)+X*Y+Z) = \pi(i(H)\varphi') = H\varphi$ and we have the
first statement of the theorem.

Now assume $G+X*Y$ is defined for some $X, Y \subseteq V$.
Partition $Y = Y_1 \cup \cdots \cup Y_n$ such that $G+X*Y =
G+X*Y_1\cdots*Y_n$ is a sequence of elementary pivots on $G+X$.
By Lemma~\ref{lem:proj_inj}, $\pi(G+X*Y) =
\pi(G+X*Y_1\cdots*Y_n) = \pi(G)*Y_1\cdots*Y_n$. By replacing
each edge complementation $*Y_i$ with $Y_i = \{u_i,v_i\}$ by
either sequence $*\{u_i\}*\{v_i\}*\{u_i\}$ or sequence
$*\{v_i\}*\{u_i\}*\{v_i\}$, see
Proposition~\ref{prop:class_uvu_vuv}, we have a desired
sequence $\varphi$ of local complementations applicable to
$\pi(G)$ with $\pi(G)\varphi = \pi(G+X*Y)$.
\end{Proof}
\fi


\section{Discussion}
We have considered loop complementation ${}+X$, pivot ${}*X$, and dual
pivot $\dual X$ on both set systems and graphs, and have shown that
they can be seen as elements of order $2$ in the permutation group
$S_3$. This group structure, in addition to the commutation property
in Lemma~\ref{lem:matrix_matroid_ops}, leads to the identity $({}+X*X)^3 =
\mathrm{id}$, cf. Theorem~\ref{thm:main_result_matroids}, and to a
normal form w.r.t.~sequences of pivots and loop complementation, cf.
Theorem~\ref{thm:normal_form_pivot_loopc_matroids}.

Although the three operations are equivalent as elements of $S_3$,
they are quite different for set systems and graphs. Indeed, for set
systems, the definition of pivot is much less involved than the
(symmetrical) definitions of loop complementation and dual pivot. In
contrast, for graphs, the definition of loop complementation is much
less involved than the (symmetrical) definitions of pivot and dual
pivot. As a direct consequence of the definitions of loop complementation
and dual pivot on set systems we notice that these operations retain
the minimal and maximal elements, respectively, of the set system.

Moreover, we obtain as a special case ``modulo loops'' a
classic relation involving local and edge complementation on
simple graphs, cf. Proposition~\ref{prop:class_uvu_vuv}. Other
relations may be easily deduced, cf.
Corollary~\ref{cor:triangle_id_simple_graphs}.

Since the notions of binary matrix and set system are
incomparable w.r.t. $\mathcal{M}_{(\cdot)}$, the
operations of pivot and loop complementation for binary
matrices and set systems are also incomparable. It remains open
whether or not one may combine and generalize the two notions
and its operations of pivot and loop complementation in a
natural way.

\subsection*{Acknowledgements}
We thank the referees for their valuable comments on the paper,
and in particular for bringing
\cite{DBLP:journals/dm/Sabidussi87} to our attention. R.B. is
supported by the Netherlands Organization for Scientific
Research (NWO), project ``Annotated graph mining''.

\bibliography{../geneassembly}

\end{document}